\documentclass[journal,twoside,web]{ieeecolor}
\usepackage{tmi}
\usepackage{cite}
\usepackage{amsmath,amssymb,amsfonts}
\usepackage{graphicx}
\usepackage{textcomp}
\usepackage{bbm}
\usepackage{multirow}
\usepackage{lineno} 
\usepackage{hyperref}
\usepackage{algorithm,algpseudocode}
\usepackage{tablefootnote}
\usepackage{threeparttable}



\algnewcommand{\Inputs}[1]{%
  \State \textbf{INPUT:}
  \Statex \hspace*{\algorithmicindent}\parbox[t]{.8\linewidth}{\raggedright #1}
}
\algnewcommand{\Initialize}[1]{%
  \State \textbf{Initialize:}
  \Statex \hspace*{\algorithmicindent}\parbox[t]{.8\linewidth}{\raggedright #1}
}
\algnewcommand{\ALGORITHM}[1]{%
  \State \textbf{ALGORITHM:}
  \Statex \hspace*{\algorithmicindent}\parbox[t]{.8\linewidth}{\raggedright #1}
}
\algnewcommand{\OUTPUT}[1]{%
  \State \textbf{OUTPUT:}
  \Statex \hspace*{\algorithmicindent}\parbox[t]{.8\linewidth}{\raggedright #1}
}
\def\BibTeX{{\rm B\kern-.05em{\sc i\kern-.025em b}\kern-.08em
    T\kern-.1667em\lower.7ex\hbox{E}\kern-.125emX}}
\begin{document}
\title{Accelerating Magnetic Resonance $\text T_{1\rho}$ Mapping Using Simultaneously Spatial Patch-based and Parametric Group-based Low-rank Tensors (SMART)}
\author{Yuanyuan Liu, Dong Liang,\IEEEmembership{Senior Member, IEEE}, Zhuo-Xu Cui, Yuxin Yang, Chentao Cao, Qingyong Zhu, Jing Cheng, \IEEEmembership{Student Member, IEEE}, Caiyun Shi, Haifeng Wang, \IEEEmembership{Member, IEEE}, and Yanjie Zhu
\thanks{This study was supported in part by the National Key R\&D Program of China no . 2020YFA0712200, National Natural Science Foundation of China under grant nos. 62201561, 12226008, 62125111,  81971611, and 81901736, the Innovation and Technology Commission of the government of Hong Kong SAR under grant no. MRP/001/18X, the Guangdong Basic and Applied Basic Research Foundation under grant no. 2021A1515110540, China Postdoctoral Science Foundation under grant no. 2022M723302, and the Shenzhen Science and Technology Program under grant no. RCYX20210609104444089.(Corresponding author: Yanjie Zhu) }
\thanks{Yuanyuan Liu is with Paul C. Lauterbur Research Center for Biomedical Imaging, Shenzhen Institute of Advanced Technology, Chinese Academy of Sciences and National Innovation Center for Advanced Medical Devices, Shenzhen, Guangdong, China (e-mail: liuyy@siat.ac.cn). }
\thanks{Zhuo-Xu Cui, Qingyong Zhu and  Dong Liang are with Research Center for Medical AI, Shenzhen Institute of Advanced Technology, Chinese Academy of Sciences(e-mail:  $\left \{\text{zx.cui; qy.zhu; dong.liang}\right \}$@siat.ac.cn).}
\thanks{Yuxin Yang, Chentao Cao, Jing Cheng, Caiyun Shi, Haifeng Wang and Yanjie Zhu are with Paul C. Lauterbur Research Center for Biomedical Imaging, Shenzhen Institute of Advanced Technology, Chinese Academy of Sciences (e-mail: $\left \{\text{yx.yang; ct.cao; jing.cheng; cy.shi; hf.wang1; yj.zhu}\right \}$@siat.ac.cn).}
\thanks{Yuanyuan Liu and Dong Liang contributed equally to this study.}}

\maketitle
\begin{abstract}
Quantitative magnetic resonance (MR) $\text T_{1\rho}$ mapping is a promising approach for characterizing intrinsic tissue-dependent information. However, long scan time significantly hinders its widespread applications. Recently, low-rank tensor models have been employed and demonstrated exemplary performance in accelerating MR $\text T_{1\rho}$ mapping. This study proposes a novel method that uses spatial patch-based and parametric group-based low-rank tensors simultaneously (SMART) to reconstruct images from highly undersampled k-space data. The spatial patch-based low-rank tensor exploits the high local and nonlocal redundancies and similarities between the contrast images in $\text T_{1\rho}$ mapping. The parametric group-based low-rank tensor, which integrates similar exponential behavior of the image signals, is jointly used to enforce multidimensional low-rankness in the reconstruction process. In vivo brain datasets were used to demonstrate the validity of the proposed method. Experimental results demonstrated that the proposed method achieves 11.7-fold and 13.21-fold accelerations in two-dimensional and three-dimensional acquisitions, respectively, with more accurate reconstructed images and maps than several state-of-the-art methods. Prospective reconstruction results further demonstrate the capability of the SMART method in accelerating MR $\text T_{1\rho}$ imaging. 
\end{abstract}

\begin{IEEEkeywords}
$\text T_{1\rho}$  mapping, fast imaging, low-rank tensor, patch
\end{IEEEkeywords}

\section{Introduction}
\label{sec:introduction}
\IEEEPARstart{Q}{uantitative}  magnetic resonance (MR) $\text T_{1\rho}$ (the spin lattice relaxation time in the rotating reference frame)\cite{wang2015t1rho} is a promising approach for characterizing intrinsic tissue-dependent information. It has been used as a novel contrast mechanism in many biomedical applications, including the musculoskeletal system\cite{wang2015t1rho_1}, brain\cite{RN683,RN8}, liver\cite{RN22, RN23}, intervertebral discs\cite{RN20}, and cardiovascular imaging\cite{qi2022t1rho}, to assess early macromolecular changes rather than conventional morphological imaging. However, MR $\text T_{1\rho}$ mapping requires acquiring multiple images with different spin-lock times (TSLs). The resulting long scan time significantly hinders its widespread clinical application. Parallel imaging is the most used acceleration method for MR $\text T_{1\rho}$ mapping, but it has a limited acceleration factor, usually approximately 2\cite{RN14}. Compressed sensing (CS) \cite{RN570,RN568,RN1} utilizing sparsifying priors has been applied in fast MR parametric mapping to achieve  higher acceleration factors. However, the scan time of MR $\text T_{1\rho}$ mapping must be further shortened to facilitate clinical use. 
\par In recent years, the low-rankness of the matrix has achieved promising results in accelerating MR parametric mapping by exploiting the anatomical correlation between contrast images globally or  locally\cite{RN559, RN579, RN573, RN649}. Compared with conventional MR images, the contrast images in parametric mapping contain redundant information across the spatial domain and are highly correlated along the temporal dimension. As an extension of the matrix, the high-order tensor can capture the data correlation in multiple dimensions\cite{RN1299} beyond the original spatial-temporal domain. The sparse representation of multidimensional image structures can be effectively exploited using a high-order tensor\cite{RN678, RN676, RN680, RN632}. Therefore, high-order tensor models have been applied in many medical imaging applications and can accelerate high-dimensional magnetic resonance imaging (MRI), such as cardiac imaging\cite{RN5, RN632, RN631, RN1297, RN2000 }, simultaneous multiparameter mapping\cite{RN8}, and functional magnetic resonance imaging\cite{RN1296}. Most of these methods globally construct a tensor by directly using the entire multidimensional image series as a tensor. The sparsity of the tensor, which usually relies on the sparsity of the core tensor coefficients, is modeled as a regularization to enforce global low-rankness using a higher-order singular-value decomposition (SVD). However, global processing jointly treats multiple tissues of different types, which may lead to residual artifacts that can be ameliorated by local processing\cite{RN559, RN632}. Like group sparsity\cite{RN681, RN682}, a high-order tensor can be constructed in a patch-based local manner. Image patches with high correlation in the neighborhood are extracted and rearranged to form local tensors. Patch-based low-rank tensor reconstruction using the local processing approach has been demonstrated to outperform low-rank tensor reconstruction globally\cite{RN632,RN1299,RN1297,RN2000, RN1296}.
\par The signal evolution in quantitative MRI exhibits a low-rank structure in the temporal dimension that can be exploited to construct a low-rank Hankel matrix\cite{RN634, RN594} to shorten the scan time further. It differs from the spatial low-rankness that processes the entire image\cite{RN559, RN579, RN638}; and the Hankel low-rank approximation treats the signal at each spatial location independently. Recently, \textit k-space low-rank Hankel tensor completion approaches have demonstrated the suitability and potential of applying tensor modeling in the \textit k-space of parallel imaging\cite{RN645, RN694}. The block-wise Hankel tensor reconstruction method\cite{RN694} was proposed by organizing multicontrast \textit  k-space datasets into a single block-wise Hankel tensor, which exploits the multilinear low-rankness based on shareable information in datasets of identical geometry. Motivated by this block-wise Hankel tensor method, tissues with equal parametric values can also be clustered in image space. Each Hankel matrix of the same tissue can be concatenated into a Hankel tensor, which exhibits significant low rankness based on high correlations in tissues of similar signal evolution.
\par This study proposes a novel and fast MR $\text T_{1\rho}$ mapping method by exploiting low-rankness in both spatial and temporal directions. First, similar local patches were found and extracted from the images to form high-order spatial tensors using a block matching algorithm\cite{RN2001}. Second, assuming that the $\text T_{1\rho}$ values of the same tissue are similar, signals were classified into different tissue groups through the histogram of the $\text T_{1\rho}$ values. In each group, signal evolution in the temporal direction was used to construct the Hankel matrix, and all groups were cascaded to form a parametric tensor. Both the spatial and parametric tensors were utilized to reconstruct the contrast images from highly undersampled \textit k-space data through high-order tensor decomposition (simultaneous \underline {s}patial patch-based and para\underline{m}etric group-b\underline{a}sed low-\underline{r}ank \underline{t}ensor, (SMART)). We used quantitative $\text T_{1\rho}$ mapping of the brain to evaluate the performance of the proposed method. Two-dimensional (2D) and three-dimensional (3D) $\text T_{1\rho}$-weighted images were undersampled with different acceleration factors to substantiate the suitability of SMART in improving the quality of fast $\text T_{1\rho}$ mapping regarding several state-of-the-art methods.
\section{Material and Methods}
\subsection{Notation}
Scalars are denoted by italic letters (e.g., $b$). Vectors are denoted by bold lowercase letters (e.g., $\textbf b$ ). Matrices are denoted by bold capital letters (e.g., $\textbf B$). Tensors of order three or higher are denoted by Euler script letters (e.g., $\mathcal{X}$). Indexed scalar $x_{ijk}$ denotes the  $(i,j,k) $th element of third-order tensor $\mathcal{X}$. Operators are denoted by capital italic letters (e.g., $C$). $\ C^T$ is used to denote the adjoint operator of $C$.
  \par The mode-$i$ matricization (unfolding) of a tensor  $\mathcal X \in \mathbb{C}^{N_1 \times N_2\times \ldots \times  N_d}$ is defined as $\textbf X_{(i)} \in \mathbb{C}^{ N_i \times \left(N_1\ldots  N_{i-1}N_{i+1}\ldots N_d\right )}$, arranging the data along the $N_{i}$th dimension to be the columns of $\textbf X_{(i)}$. The opposite operation of tensor matricization is the folding operation, which arranges the elements of the matrix $\textbf X_{(i)}$ into the $d$th order tensor $\mathcal X$. Tensor multiplication is more complex than matrix multiplication. In this study, only the tensor $i$-mode product\cite{RN763} is considered, which is defined as the multiplication (e.g., $\mathcal X \times_i \mathbf U$) of a tensor $\mathcal X$ with a matrix $\textbf U \in \mathbb C^{J\times N_i}$ in mode $i$. Elementwise, we have $ w_{n_{1}\cdots n_{i-1} j n_{i+1}\cdots n_{d}} =\sum_{n_{i}=1}^{N_{i}} x_{n_{1} \cdots n_{i-1} n_{i}n_{i+1}\cdots n_{d}} u_{j n_{i}} $, where $\mathcal W = \left(\mathcal{X} \times{ }_{i} \mathbf{U}\right )\in \mathbb{C}^{N_1 \times \ldots \times N_{i-1}\times J \times N_{i+1}\times \ldots \times N_d}$. The $i$-mode product can also be calculated by the matrix multiplication $\mathbf W_{(i)} = \mathbf U \mathbf X_{(i)}$.

 \subsection{Measurement Model}
Let $\textbf X \in \mathbb{C}^{N_\textit{voxel} \times N_{\textit {TSL}}}$ be the $\text T_{1\rho}$-weighted image series to be reconstructed, where $\textit{N}_\textit{voxel}$ denotes the voxel number of the
$\text T_{1\rho}$-weighted image, and $\textit{N}_\textit{TSL}$ is the number of TSLs. $\mathbf Y \in \mathbb{C}^{N_\textit{voxel} \times N_c \times N_{\textit {TSL}}}$ denotes the corresponding \textit k-space measurement, where $N_c$ denotes the coil number. The forward model for $\text T_{1\rho}$ mapping is given by 
\begin{equation}
    \mathbf Y = E\mathbf X + \zeta,
\end{equation}
where $\zeta$ denotes the measurement noise and $E$ denotes the encoding operator\cite{RN636,RN637} given by $E=AFS$, where $A$ is the undersampling operator that undersamples $k$-space data for each $\text T_{1\rho}$-weighted image. $F$ denotes the Fourier transform operator. $S$ denotes an operator which multiplies the coil sensitivity map by each $\text T_{1\rho}$-weighted image coil-by-coil. The general formulation for recovering $\mathbf X$ from its undersampled measurements can be formulated as

\begin{equation}
  \begin{gathered}
\underset{\textbf X}{\operatorname{\arg \min}} \frac{1}{2}\left \|E \mathbf X-\mathbf Y \right\|_{F}^{2} +\lambda R(\mathbf X) 
\end{gathered},  
\end{equation}
where  $\left\| \cdot \right\|_{F}$ denotes the Frobenius norm, $R(\mathbf X)$  denotes a combination of 
regularizations, and $\lambda$ is the regularization parameter.
\par In this study, the SMART reconstruction method assumes 
that the $T_{1\rho}$-weighted image series $\mathbf X$ can be expressed as a high‐order 
low‐rank representation on a spatial patch-based tensor and a parametric group-based tensor (shown in Fig. \ref{fig1}). The problem in (2) can be formulated as a constrained optimization on the two high‐order low‐rank tensors.
\subsection{Proposed Method}
\subsubsection{Spatial Tensor Construction}
We use the block matching algorithm\cite{RN2004,RN2005} to extract similar anatomical patches for building a spatial patch-based tensor at each spatial location. $\mathbf X$ can be expressed as a high-order low-rank representation on a patch scale.
For a given reference patch $\textbf {B}_{\textit{i}}$ ( $\textbf{B}_{\textit{i}} \in \mathbb{C}^{N_b}$, $N_b=b\times b$ in 2D imaging and $N_b=b\times b\times b$ in 3D imaging ) centered at spatial location $i$, let $P_i$ denote the patch selection operator that extracts a group of similar patches to $\textbf B_{\textit i} $ 
 from all time points and constructs a low-rank tensor
$\mathcal{T}_i \in \mathbb{C}^{N_b\times N^i_{p}\times N_\textit{TSL}}$ from them, where $N^i_{p}$ denotes the number of similar patches. The process can be expressed as $\mathcal{T}_{\textit{i}}=\textit{P}_\textit{i}(\textbf{X})$, and the adjoint operation $\textit{P}_\textit{i}^\textit{T}$ places the patches back to their original spatial locations in the image.
\par The patch selection process is as follows. For the patch $\textbf{B}_\textit{i}$, it is compared with the other image patch $\textbf B_{\tilde{i}}$ based on the normalized $l_2$-norm distance \cite{RN2001} with the exact formulation of the normalized mean-squared error as
\begin{equation}
\operatorname{dist}\left(\textbf B_{i}, \textbf B_{\tilde{i}}\right)=\left\|\textbf B_{i}-\textbf B_{\tilde{i}}\right\|_{2}^{2} /\left\|\textbf B_{\tilde{i}}\right\|_{2}^{2}.
\end{equation}
 $\textbf B_{\tilde{i}}$ is considered similar to $\textbf B_i$ when the distance is less than a predefined threshold $\lambda_m$. We compare the candidates within a specified search radius $r$ and limit the maximum similar patch number to $N_{p,max}$ to reduce the complexity. If more image patches are matched to $\textbf B_i$, only the $N_{p,max}$ patches with the highest degree of similarity are involved in the group for further processing. 
\subsubsection{Parametric Tensor Construction}
Similar to $\text T_2$ mapping in previous studies \cite{RN634}, the $\text T_{1\rho}$-weighted images $I({\textit{c}},\textit{t}_m)$ in $\text T_{1\rho}$ mapping can be expressed as follows:
\begin{equation}
I\left({\textit c}, \textit {t}_{m}\right)=\sum_{l=1}^{L} \rho_{l}(c) \exp [i \varphi(\textit c)] \exp \left[-\textit{t}_{m} / T_{1 \rho,l}(\textit c)\right],
\end{equation}
where $\rho_l(\textit{c})$ represents the proton density distribution, $\text T_{1\rho, l}$ is the $\text T_{1\rho}$ relaxation value of the $l$th tissue component, $\varphi (\textit{c})$ represents the phase distribution, ${\textit{c}}$ indicates the spatial coordinate, $\textit{t}_m$ is the $m$th TSL, and $L$ is the number of linearly combined exponential terms, which is tissue-dependent.
For instance, in the three-pool model of white matter, the white matter tissue is composed of a myelin water pool, myelinated axon water pool, and mixed water pool, yielding $L=3$. 

The signals from each voxel along the temporal direction can be used to form a Hankel matrix for $\forall{{\textit{c}}}\in {\boldsymbol{\Omega}}$ (where $\boldsymbol{\Omega}$ denotes the spatial support of the tissue.)
\begin{equation}
\label{Hankel matrix}
\setlength{\arraycolsep}{1pt}
 \begin{array}{lc}
\mathbf G[I({\textit{c}})]= \begin{bmatrix}
 I({c},t_1)& I({c},t_2)& \cdots& I({c},t_k) \\ 
 I({c},t_2)& I({c},t_3)& \cdots & I({c},t_{k+1}) \\ 
 \vdots & \vdots &\vdots& \vdots  \\ 
 I({c},t_{N_{\textit{TSL}}-k+1})& I({c},t_{N_{\textit{TSL}}-k+2})& \cdots & I({c},t_{N_{\textit{TSL}}}) \end{bmatrix}
\end{array}
 \end{equation}
with a low rank property (e.g., $\text {rank}(\mathbf G)= L 
 (L\ll N_\textit{TSL})$, where $\text{rank}(\mathbf G)$ denotes the rank of matrix $\mathbf G$, and the theoretical proof can be found in \cite{RN634}). The Hankel matrix is designed as square as possible, where $k$ is selected as the nearest integer greater than or equal to half the number of TSLs.
\par The parametric tensor is constructed using the tissue clustering method as follows: First, a $\rm{T_{1\rho}}$ map is estimated from the initial $\rm T_{1\rho}$-weighted image reconstructed using zero-filling. A histogram analysis divides the $\rm T_{1\rho}$ map into $ N_g$ groups. Each group is considered a group of signals belonging to one tissue category. According to linear predictability\cite{RN634, RN686}, the signals from each voxel along the temporal direction can form a Hankel matrix. The Hankel matrices belonging to the same group are combined as a tensor $\mathcal{Z}_j \in \mathbb{C}^{N^j_{\textit{tissue}}\times{ (N_{\textit{TSL}}-k+1)}\times k}$ from $\textbf X$. This can be expressed as $\mathcal{Z}_{j}=H_{j}(\textbf X)$, ($j=[1,2,\cdots,N_g]$), where $N^j_\textit{tissue}$ denotes the number of pixels in each group, and $H$ denotes the parametric tensor construction operator.

\begin{figure*}[!t]
\centering{\includegraphics[width=2\columnwidth]{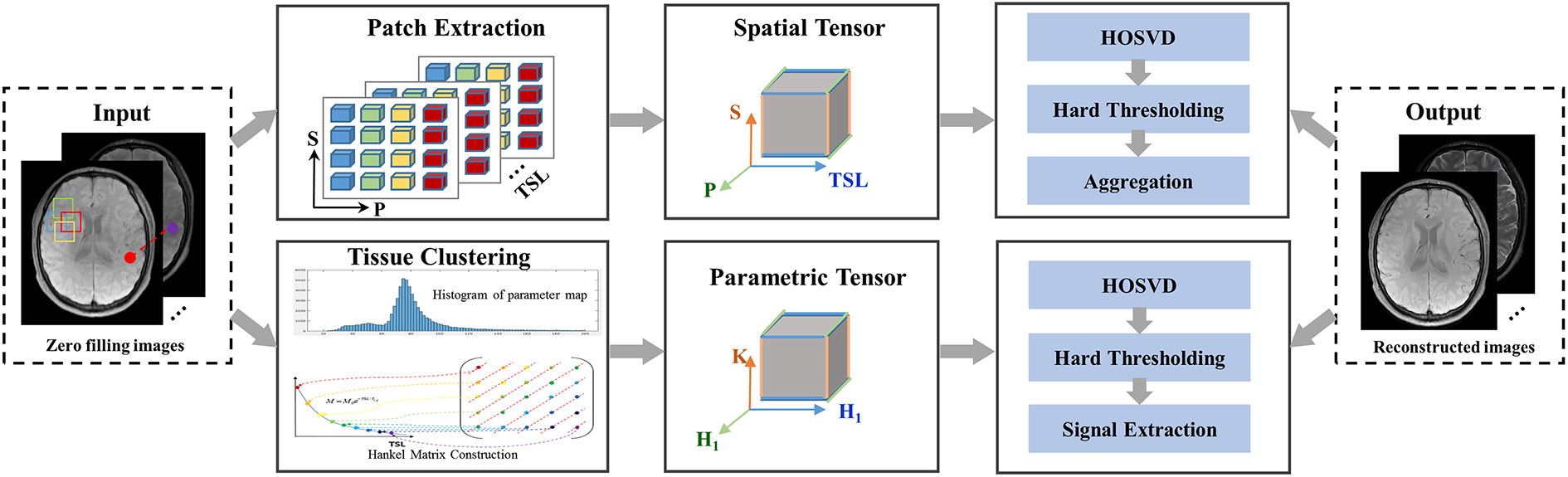}}
\caption{Flow diagram of the proposed SMART method’s optimization for subproblem (P2) and subproblem (P3). Groups of similar 2D (respectively 3D) patches from the $\text T_{1\rho}$-weighted images are extracted using the block matching method to construct spatial tensor $\mathcal{T}_i$. A third-order tensor can  be formed by stacking each vectorized similar patch along the group and TSL direction. The $\text T_{1\rho}$ values are estimated from the $\text T_{1\rho}$-weighted images using a nonlinear fitting method according to the signal relaxation model to construct the parametric tensor $\mathcal{Z}_j$. The image can be grouped into several groups of tissues after applying histogram analysis of the $\text T_{1\rho}$. Moreover, in each group of the same tissue, the vector of pixels at a fixed location in the contrast images along the TSL direction can be extended to a Hankel matrix. In each group, all the Hankel matrices for the signals can be constructed as a third-order tensor.}
\label{fig1}
\end{figure*}

\subsubsection{SMART Reconstruction Model}
Based on the spatial and parametric tensors, we propose a novel method that simultaneously uses spatial patch-based and parametric group-based low-rank tensor through higher-order tensor decomposition. Let $\mathcal{T}=[\mathcal T_1, \mathcal T_2,\ldots,\mathcal T_i,\ldots]$, and  $\mathcal{Z}=[ \mathcal Z_1, \mathcal Z_2,\ldots, \mathcal Z_j, \ldots]$. $\rm T_{1\rho}$-weighted images can be reconstructed from the undersampled $\textit k$-space data by solving the following optimization problem:
\begin{equation}
\label{SMART Model}
\begin{gathered}
\underset{\mathbf X}{\arg \min} \frac{1}{2}\left \|E \textbf X-\textbf Y \right\|_{F}^{2}+\lambda_{1}\sum_{i} \left\|\mathcal T_{i}\right\|_{*}+\lambda_{2}\sum_{j} \left\|\mathcal Z_{j}\right\|_{*} \\
\text { s.t. } \mathcal{T}_{i}=P_{i}(\textbf X), \mathcal Z_{j}= H_{j}(\textbf X),
\end{gathered}
\end{equation}
where $\left\| \cdot \right\|_{*}$ is the nuclear norm, $\lambda_{1}$ and $\lambda_{2}$ are the regularization parameters.
As the operator $P_i$ is time-varying while the operator $H_j$ always extracts mask on the same locations, (\ref{SMART Model}) can be transformed into the following formulation using the Lagrangian optimization scheme:
\begin{equation}
\label{Lagrangian model}
\begin{gathered}
L^n\left(\textbf X, \mathcal{T}, \mathcal{Z}, \alpha_{1}, \alpha_{2}\right):=\frac{1}{2}\left\|E \textbf X-\textbf Y\right\|_{F}^{2}+\lambda_{1}\sum_{i}
\left\|\mathcal{T}_{i}\right\|_{*}+\\
\frac{\mu_{1}}{2} \sum_{i}\left\| \mathcal{T}_{i}-P^n_{i}(\textbf X)\right\|_{F}^{2}-\sum_{i}\left\langle \alpha_{1i}, \mathcal{T}_{i}-P^n_{i}(\textbf X)\right\rangle+ 
\lambda_{2}\sum_{j} \left\|\mathcal{Z}_{j}\right\|_{*}\\+\frac{\mu_{2}}{2} \sum_{j} \| \mathcal{Z}_{j}-{H}_{j}(\textbf X)\|_{F} ^{2}-\sum_{j} \left\langle\alpha_{2j}, \mathcal{Z}_{j}-{H}_{j}(\textbf X)\right\rangle,
\end{gathered}
\end{equation}
where $\mu_1$ and $\mu_2$ denote the penalty parameters, $n$ denotes the $n$th iteration number, and $\alpha_1=[\alpha_{11},\alpha_{12},\ldots,\alpha_{1i},\ldots]$ and $\alpha_2=[\alpha_{21},\alpha_{22},\ldots,\alpha_{2j},\ldots]$ are the Lagrange multipliers. For simplicity, (\ref{Lagrangian model}) can be rewritten as follows: 
\begin{equation}
\label{SMART Model with Lagrangian}
\begin{gathered}
L^n\left(\textbf X, \mathcal{T}, \mathcal{Z}, \alpha_{1}, \alpha_{2}\right):=\frac{1}{2}\left\|E \textbf X-\textbf Y\right\|_{F}^{2}+\lambda_{1}\sum_{i}
\left\|\mathcal{T}_{i}\right\|_{*}+\\
\frac{\mu_{1}}{2} \sum_{i}\| \mathcal{T}_{i}-P^n_{i}(\textbf X)-\frac{\alpha_{1i}}{\mu_1}\|_{F}^{2}+ 
\lambda_{2}\sum_{j} \left\|\mathcal{Z}_{j}\right\|_{*}+\\
\frac{\mu_{2}}{2} \sum_{j} \| \mathcal{Z}_{j}- {H}_{j}(\textbf X)-\frac{\alpha_{2j}}{\mu_2}\|_{F} ^{2} .
\end{gathered}
\end{equation}
\par Equation (\ref{SMART Model with Lagrangian}) can be efficiently solved through operator splitting via the alternating direction method of multipliers (ADMM)\cite{RN695} by decoupling the optimization problem into three subproblems:\\
\underline{Update on $\textbf X$ (subproblem P1):}
\begin{equation}
\begin{gathered}
L^n_1\left( \textbf X\right)
:=\underset{\mathbf {X}}{\arg \min}
\frac{1}{2}\left\|E \textbf X-\textbf Y\right\|_{F}^{2}+\\
\frac{\mu_{1}}{2} \sum_{i}\| \mathcal{T}_{i}-P^n_{i}(\textbf X)-\frac{\alpha_{1i}}{\mu_1}\|_{F}^{2} 
+\frac{\mu_{2}}{2} \sum_{j} \| \mathcal{Z}_{j}-{H}_{j}(\textbf X)-\frac{\alpha_{2j}}{\mu_2}\|_{F} ^{2}.
\end{gathered}
\end{equation}
The solution \textbf X for problem (P1) can be effectively solved using the conjugate gradient (CG) algorithm\cite{RN2002}.
\underline{Update on $\mathcal{T}$ (subproblem P2):}
\begin{equation}
\begin{gathered}
L^n_2\left(\mathcal{T} \right):=\underset{\mathcal{T}}{\arg \min}
 \ \lambda_{1}\sum_{i}
\left\|\mathcal{T}_{i}\right\|_{*}+\\
\frac{\mu_{1}}{2} \sum_{i}\| \mathcal{T}_{i}-P^n_{i}(\textbf X)-\frac{\alpha_{1i}}{\mu_1}\|_{F}^{2}.
\end{gathered}
\end{equation}
Subproblem (P2) can be solved in three steps. First, higher-order singular-value decomposition (HOSVD) is applied to the tensor $\mathcal{T}_i$\cite{RN632, RN645, RN644}. It decomposes  $\mathcal{T}_i$ into a core tensor $\mathcal{G}$ and three orthonormal bases $\textbf U^{(e)}( e= 1,2,3)$ for the three different subspaces of the $e$-mode vectors\cite{RN763}:
\begin{equation}
\mathcal{T}_{i}=\mathcal{G} \times_{1} \textbf U^{(1)} \times_{2} \textbf U^{(2)} \times_{3} \textbf U^{(3)},
\end{equation}
where $\textbf U^{(e)}$ is an orthogonal unitary matrix  obtained from the SVD of $\textbf T_{i(e)}$. The low-rank tensor approximation is typically performed by soft thresholding\cite{cai2010singular} the core tensor or truncating the core tensor and unitary matrices when the $e$-mode ranks are known \cite{RN1296}. Moreover, while the use of hard thresholding only provides an approximate solution to the nuclear norm regularized optimization problem \cite{RN644,RN633,RN1301}, we found that both solutions achieved similar performance in the proposed SMART algorithm.
As one of the comparison methods (PROST) \cite{RN633} in this study used hard thresholding in HOSVD, we also used hard thresholding for a fair comparison.

The low-rank tensor approximation effectively acts as a high-order denoising process, where the small discarded coefficients mainly reflect contributions from noise\cite{RN644} and noise-like artifacts\cite{RN645,RN633}. An implementation of the HOSVD is shown in Algorithm 1 of the supplementary information.  Second, the denoised $\mathcal{T}_i$ tensors were rearranged to form the denoised patches. Finally, the image patches overlap can be combined by simple averaging (Fig. \ref{fig1}, “Aggregation”) to generate an estimated image.
\par \underline{Update on $\mathcal{Z}$ (subproblem P3):}
\begin{equation}
\begin{gathered}
L^n_3\left(\mathcal{Z} \right):=\underset{\mathcal{Z}}{\arg \min} \ \lambda_{2}\sum_{j}
\left\|\mathcal{Z}_{j}\right\|_{*}+\\
\frac{\mu_{2}}{2} \sum_{j}\| \mathcal{Z}_{j}-H_{j}(\textbf X)-\frac{\alpha_{2j}}{\mu_2}\|_{F}^{2}.
\end{gathered}
\end{equation}
Subproblem (P3) can also be solved in three steps. First, perform low-rank tensor approximation similar to the first step in solving subproblem (P2). Second, the image signals were extracted from the Hankel matrix from each horizontal slice of $\mathcal{Z}_j$ along the first dimension. Finally, an estimated image can be generated after repeating the above two steps for all parametric tensors. 
\par \underline{Update on $\alpha_1$ and $\alpha_2$:}
\begin{equation}
    \alpha_{1i}^{n+1}=\alpha_{1i}^n+\mu_1[P_i^{n+1} (\textbf X^{n+1})-\mathcal{T}_i^{n+1}],
\end{equation}
\begin{equation}
    \alpha_{2j}^{n+1}=\alpha_{2j}^n+\mu_2[H_j (\textbf X^{n+1})-\mathcal{Z}_j^{n+1}].
\end{equation}
\par An implementation of the SMART method is shown in Algorithm 2 of the supplementary information.

\subsubsection{Space complexity analysis}
The proposed SMART method consists of four components: solving the three sub-optimization problems P1, P2, and P3, and updating $\alpha_1$ and $\alpha_2$. We analyze the above components to investigate their space complexity using the $\mathcal{O}$ notation. Assume that the number of spatial tensors is $N_\mathcal{T}$, and the number of parametric tensors is $ N_\mathcal{Z}$. Solving $\mathbf X$ subproblem (P1) is dominated by the CG algorithm with cost of $\mathcal{O}((N_\textit{voxel}\times N_\textit{TSL})^{4/3})$\cite{RN2010}. Solving $\mathcal T$ of subproblem (P2) is dominated by  the HOSVD operation, which applies SVD to each mode of the spatial tensor, which costs $\mathcal{O}(N_\mathcal{T}\times [{N_b}\times (N_p^i \times N_\textit {TSL})^2+{N_p^i}\times (N_b \times N_\textit{TSL})^2 +{N_\textit{TSL}}\times (N_b \times N_p^i)^2])$. Similarly, solving $\mathcal{Z}$ of subproblem (P3) costs $\mathcal{O} ( N_{\mathcal{Z}} \times[N_{\text {tissue }}^{j} \times((N_{\textit {TSL }}-k+1) \times k)^{2}+(N_{\textit {TSL }}-k+1) \times (N_{\text {tissue}}^{j} \times k)^{2}+k \times (N_{\text {tissue }}^{j} \times(N_{\textit {TSL }}-k+1))^{2}])$. Updating $\alpha_1$ and $\alpha_2$ costs $\mathcal{O}(N_\textit{voxel}\times N_\textit{TSL} )$. Therefore, the asymptotic space complexity of the SMART method is the upper bound of the space complexity of the above components.

\subsection{Parameter Selection}
The SMART method performance is affected by the parameters of the spatial and parametric tensor construction.These include patch size $\textit b$, maximum similar patch number $ N_{p,max}$, normalized $l_2$-norm distance threshold $\lambda_m$ in the spatial tensor construction, and the group number $\textit N_{g}$ in the parametric tensor construction. These parameters should be carefully tuned to obtain the best reconstruction performance. 

\subsection{$\text T_{1\rho}$ Map Estimation}
$\rm T_{1\rho}$ maps were obtained by fitting the reconstructed $\rm T_{1\rho}$-weighted images with different TSLs pixel-by-pixel:
\begin{equation}
\label{momo_expo}
M_{k}=M_{0} \exp \left(-t_k/ T_{1 \rho}\right)_{k=1,2, \ldots, N_{\textit {TSL }}},
\end{equation}
where $M_0$ denotes the baseline image intensity without applying a spin-lock pulse, and $M_k$ is the signal intensity for the \textit{k}th TSL image. $\text T_{1\rho}$ map was estimated using the nonlinear least-squares fitting method with the Levenberg–Marquardt algorithm\cite{RN1179} from the reconstructed $\text T_{1\rho}$-weighted images.

\section{EXPERIMENT}
\subsection{Evaluation of the Tissue Clustering}
 Numerical phantom, real phantom, and in vivo experiments were performed to evaluate the tissue clustering method performance. All MR scans involved were performed on 3T scanners. The details of data acquisition are shown in the supplementary information and Table S1. 
 \subsubsection{Numerical phantom dataset}The numerical phantom consisted of five vials (including 2809 pixels in each vial) with different  $\text T_{1\rho}$ values representing five brain regions: putamen, frontal white matter, genu corpus callosum, head of caudate nucleus, and centrum semiovale. The numerical phantom was simulated using a bi-exponential model  \cite{RN688,RN17} with the following formula to mimic multi-component tissue:
\begin{equation}
\label{bi_expo}
\begin{gathered}
M_{k}=M_{0}((1-\alpha) \exp (-t_{k} / T_{1 \rho ,\textit {short}})+\\
\alpha \exp (-t_{k} / T_{1 \rho, \textit {long}}))_{k=1,2, \ldots, N_\textit{TSL}},
\end{gathered}
\end{equation}
where  $T_{1 \rho, \textit {long}}$ and  $T_{1 \rho, \textit{ short}}$ are the long and short  $\text T_{1\rho}$ relaxation times, $\alpha (0\leq\alpha\leq1)$ and $1-\alpha$ are the fractions for  these two components, respectively. For the five vials, the $T_{1 \rho, \textit{long}}$ values were set as 77, 78, 79, 82, and 89 ms. The $T_{1 \rho, \textit{short}}$ values were set as 18, 19, 20, 21, 22 ms, and $\alpha$ was set as 0.6, according to the previous study\cite{2017biexpo}. The datasets were simulated with the matrix size = $192\times 192$, and TSLs = 1, 20, 40, 60, and 80 ms.
\subsubsection{Real phantom dataset} The phantom contained nine $\text{NiCl}_2$-doped agarose vials, with  $\text{T}_1$ and $\text{T}_2$ values in the in vivo range ($\text{T}_1$: 250 ms-1872 ms; $\text{T}_2$: 42 ms-231 ms)\cite{T1T2phantom}.  $\text T_{1\rho}$-weighted images with TSLs = 1, 20, 40, 60, and 80 ms were  acquired with ten averages, leading to a high signal-to-noise ratio (SNR) of the images. The exponential model of (\ref{momo_expo}) was applied to obtain the $\text T_{1\rho}$ map. 
 Supplementary information Fig. S1(a-c) shows a $\text T_{1\rho}$-weighted image of the phantom at TSL = 40 ms, the  $\text T_{1\rho}$ map, and the estimated $\text T_{1\rho}$ values of the nine vials, respectively.

\subsubsection{In vivo dataset} One 2D brain dataset was collected from a volunteer with the same TSLs used in the phantom experiment. The average number was set as one, considering the long scan time.
\par Complex Gaussian noise was added to the numerical, real phantom, and in vivo datasets with SNRs ranging from 25 to 60 with an increment of five. SNR was computed as the ratio between the average value of $\text T_{1\rho}$-weighted images and the standard deviation of the noise \cite{RN573}. 
Two rank values were calculated for the images with different SNRs. One group consisted of the mean rank of all the Hankel matrices for each pixel in the region of interest (ROI).The other was the rank of a block Hankel matrix for a specific tissue. Let $\mathbf{G}_i$ denote the Hankel matrix constructed from the pixels at a fixed location $i$ in the clustered tissue  according to (\ref{Hankel matrix}) (shown in Fig.\ref{fig1}). The block Hankel matrix can be defined as $ \mathbf G_\text{Block}=\left[\mathbf G_{1}, \mathbf G_{2}, \ldots, \mathbf G_{N_v}\right]$, where $N_v$  denotes the total number of pixels in the clustered tissue.

\par 
The rank value was calculated using the SVD with a fixed singular-value threshold calculated as the product of an empirical ratio and the most significant singular value. 
The ratio was set as 0.03 for both groups. The Monte Carlo simulation was performed 1000 times for each SNR to reduce the noise bias on the rank value calculated.

\par The $k$-space data were undersampled with acceleration factor (R) R = 4 for the numerical and real phantom datasets, and R = 6 for the in vivo dataset to analyze the effect of residual aliasing. Initial reconstructions of the undersampled data were obtained using the Sparse MRI method\cite{RN568} and were used to calculate the rank value.

\subsection{In Vivo Reconstruction Experiments}
The proposed method was evaluated on the $\text T_{1\rho}$ mapping datasets of the brain collected from six volunteers on 3T scanners\cite{RN680, RN687}. The local institutional review board approved the experiments, and informed consent was obtained from each volunteer. The details of data acquisition are shown in supplementary information and Table S1. 

\par For the 2D datasets, the fully sampled \textit k-space data were retrospectively (Retro) undersampled along the ky dimension using a pseudo-random undersampling pattern \cite{RN646} with acceleration factors (R) = 4 and 6. For the 3D datasets, the fully sampled \textit k-space datasets were retrospectively undersampled using Poisson disk random\cite{RN16} patterns with R = 6.76 and 9.04. The sampling masks for each TSL were different. For the prospective (Pro) study , two 2D datasets were acquired from one volunteer (27 years old) with R = 4.48 and 5.76, respectively. Fully sampled data were also acquired as a reference in the prospective experiment.

\par SMART and four state-of-the-art methods were used to reconstruct the undersampled \textit k-space data. These methods include the high-dimensionality undersampled patch-based reconstruction method (PROST) \cite{RN633} using the spatial patch-based low-rank tensor, low-rank plus sparse (L + S) method\cite{RN636} enforcing the spatial global low-rank and sparsity of the image matrix, locally low-rank method (LLR)\cite{RN559} using the spatial patch-based low-rank matrix, and model-driven low rank and sparsity priors method (MORASA)\cite{RN634} which enforces the global low-rankness of the matrix in both spatial and temporal directions and sparsity of the image matrix. In addition, a modified PROST reconstruction method (PROST + HM) was also compared with SMART to verify the effectiveness of the parametric group-based low-rank tensor in image reconstruction. This jointly uses the spatial patch-based low-rank tensor and a parametric low-rank matrix by replacing the patch-based parameter tensor in SMART with a Hankel matrix. Experiments with high acceleration factors of 10.2 and 11.7 in the 2D scenario and 11.26 and 13.21 in the 3D scenario were also performed to verify the effectiveness of the proposed method.

\begin{figure*}[!htbp]
\centering{\includegraphics[width=2\columnwidth]{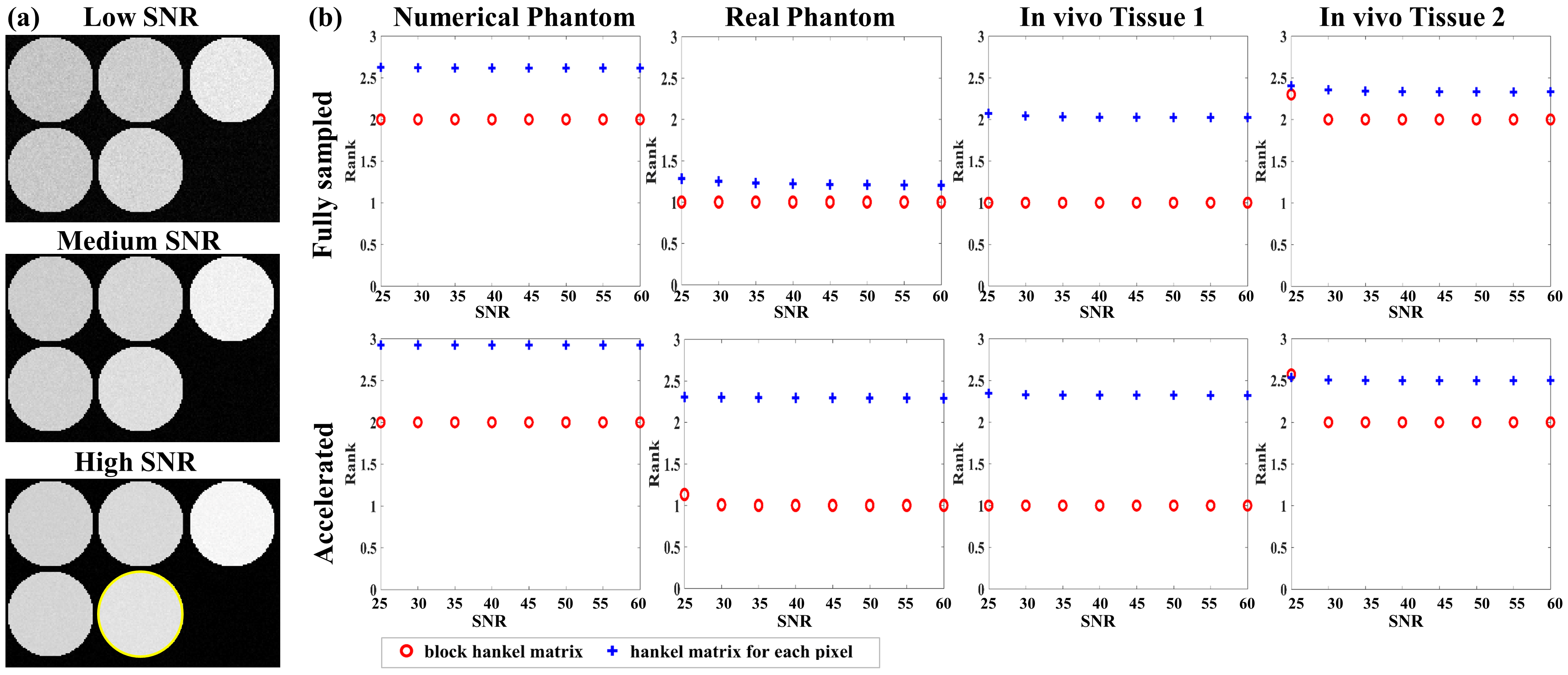}}
\caption{ Mean rank of all the Hankel matrices and the rank of the block Hankel matrix for each pixel  in selected ROIs. (a) The $\text T_{1\rho}$-weighted images of the numerical phantom at TSL = 80 ms, with SNR of 25 (low), 40 (medium), and 60 (high).The vial marked with a yellow circle shows the selected ROI for the rank calculation of the numerical phantom. (b) Rank curves as a function of SNR for the fully sampled data, and the accelerated data, respectively (from left to right:  numerical phantom data, real phantom data, in vivo tissue 1, in vivo tissue 2).}
\label{fig2}
\end{figure*}

\par $\text T_{1\rho}$ quantification was assessed using the normalized root mean square error (nRMSE)\cite{RN688}. The quality of the reconstructed $\text T_{1\rho}$-weighted images was assessed based on the nRMSE, high-frequency error norm (HFEN)\cite{RN689,RN648}, structural similarity index measure (SSIM)\cite{RN647}, and peak signal-to-noise ratio (PSNR).
\subsection{Parameter Selection}
In this subsection, several different values of the parameters related to the spatial and parametric tensor construction mentioned in the previous section were set and used for 2D image reconstruction. The nRMSE values for the reconstructions of SMART with R = 4 and R = 6 using different parameters were compared.
\par The representative zero-filling reconstruction of the Retro 2D dataset with R = 6, the intermediate reconstruction at the 12th iteration, the parameter histogram, the estimated $\text T_{1\rho}$ map, the corresponding tissue groups, and the ranks of tissue groups were also compared to demonstrate the necessity of updating tissue maps in the iteration.

All reconstructions, $\text T_{1\rho}$ estimation, and analyses were performed in MATLAB 2017b (MathWorks, Natick, MA, USA) on an HP workstation with 500 GB DDR4 RAM and 32 cores (two 16-core Intel Xeon E5-2660 2.6 GHz processors). The reconstruction parameters used in all methods are presented in the supplementary information Table S2. The tensor transforms, including the folding and unfolding operators, and the tensor multiplication were implemented using the MATLAB tensor toolbox by Brett \textit{et al.} from Sandia National Laboratories\footnote{\url{http://www.tensortoolbox.org/}}. The HOSVD using algorithm 1 of the supplementary information was implemented based on the transforms from the above toolbox.

\begin{figure*}[!htbp]
\centering{\includegraphics[width=1.9\columnwidth]{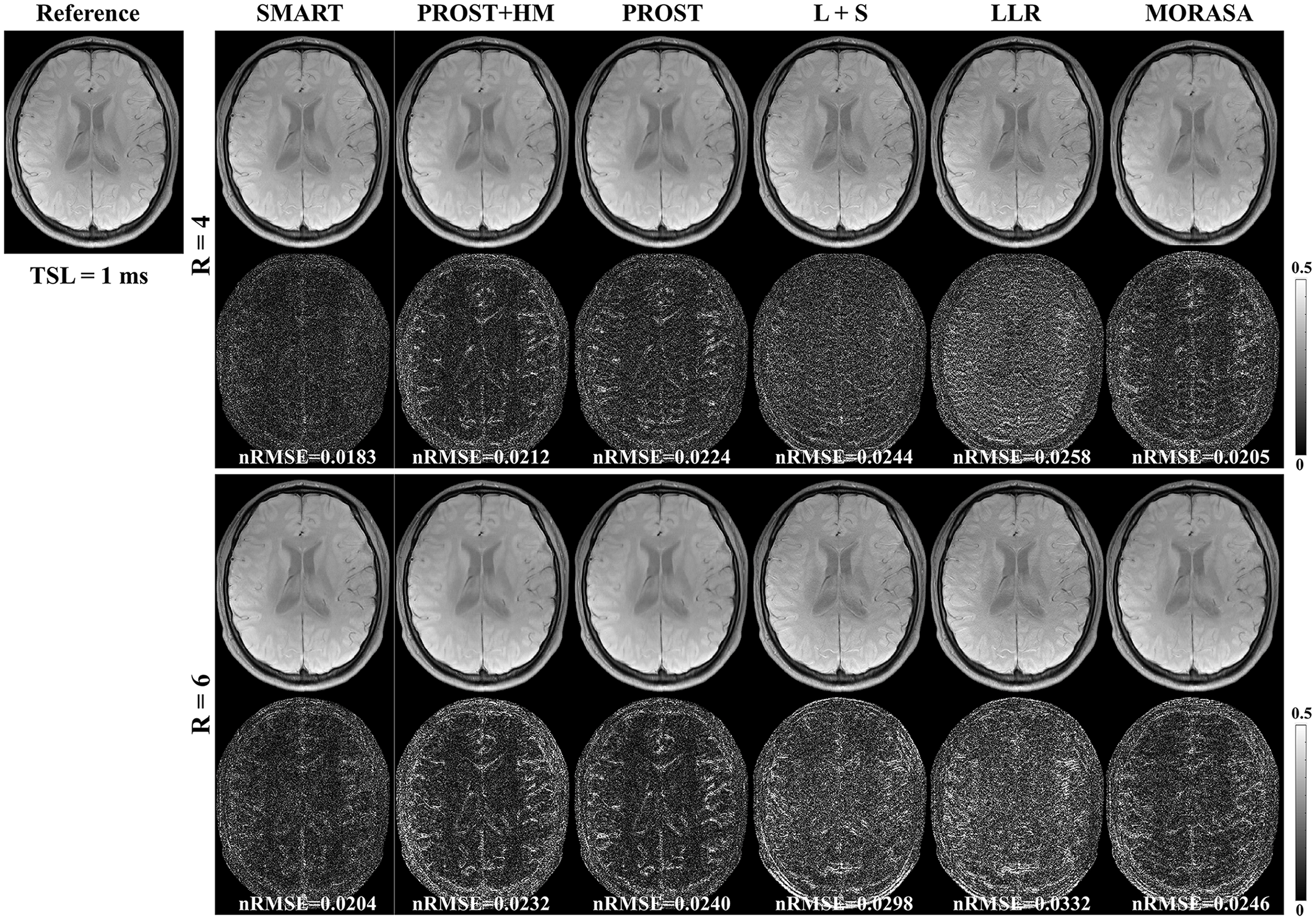}}
\caption{Reconstructed $\text T_{1\rho}$-weighted images from one 2D brain dataset at TSL = 1 ms with acceleration factors (R) = 4 and 6 using the SMART, PROST+HM, PROST, L+S, LLR, and MORASA methods. The corresponding error images for the reference image reconstructed from the fully sampled data are also shown. The error images are amplified by ten for visualization. The nRMSEs are shown at the bottom of the error images.}
\label{fig4}
\end{figure*}

\subsection{Ablation Experiment }
We conducted an ablation study to analyze the spatial and parametric tensors’ effects on the reconstruction. Two models are constructed as follows:\\
\underline{model 1:}
\begin{equation}
\underset{\mathbf X}{\arg \min } \frac{1}{2}\|E \mathbf X-\mathbf Y\|_{F}^{2}+\lambda_{1}\sum_{i} \left\|\mathcal{T}_{i}\right\|_{*} \text { s.t. } \mathcal{T}_{i}=P_{i}(\textbf X)
\end{equation}
\underline{model 2:}
\begin{equation}
\underset{\mathbf X}{\arg \min } \frac{1}{2}\|E \textbf X-\mathbf Y\|_{F}^{2}+ \lambda_{2}\sum_{j} \left\|\mathcal{Z}_{j}\right\|_{*} \text { s.t. } \mathcal{Z}_{j}={H}_{j}(\mathbf X)
\end{equation}
 Model 1 only includes the spatial tensor, and model 2 only includes the parametric tensor. Model 1 is the same as the PROST method. Similarly, we applied ADMM to solve the optimization problem in model 2. 
These two models were used to reconstruct images from the retrospectively undersampled in vivo dataset with R = 4 and 6.


\begin{figure*}[!htbp]
\centerline{\includegraphics[width=17cm]{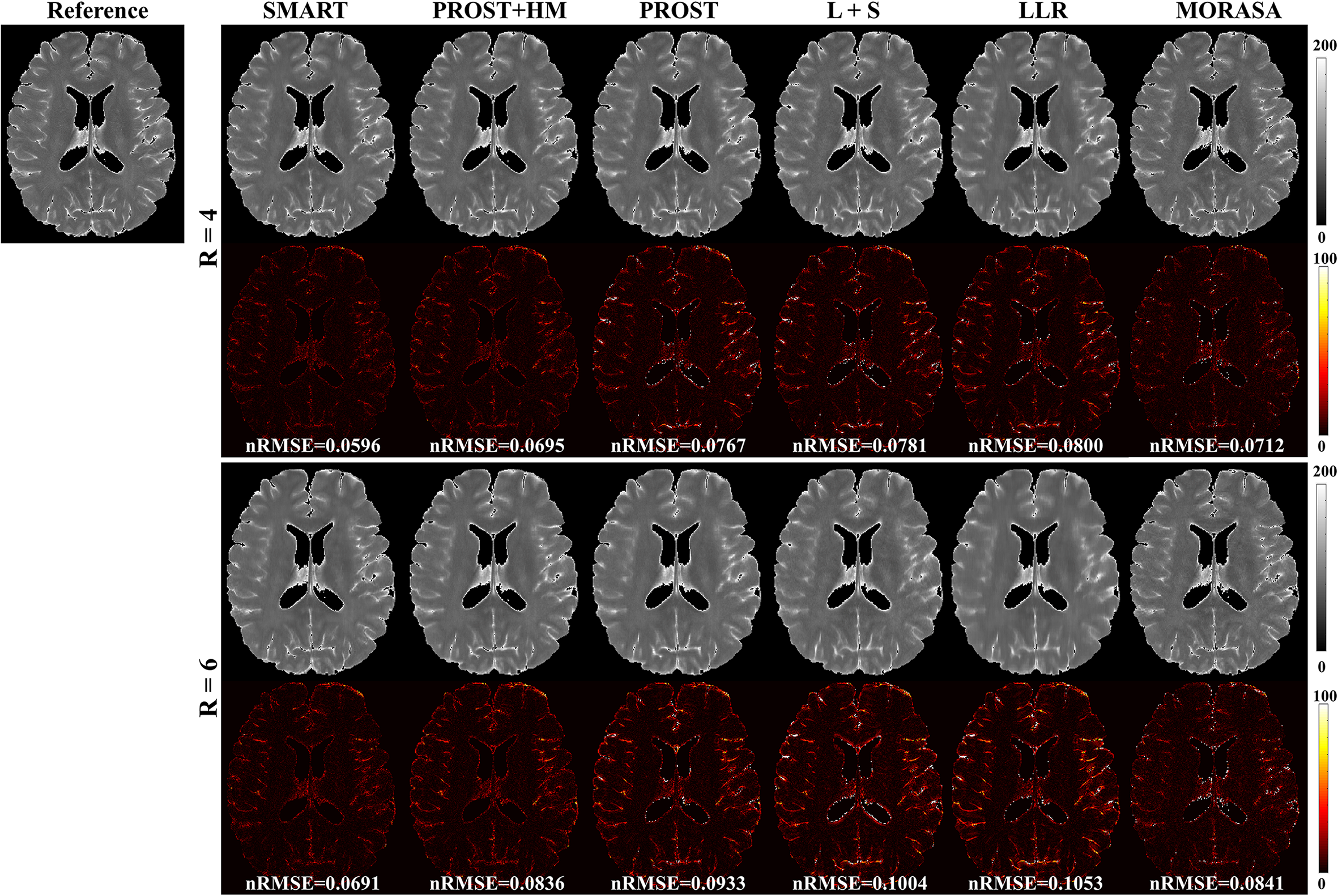}}
\caption{ Reconstructed $\text T_{1\rho}$maps estimated from one 2D brain dataset with acceleration factors (R) = 4 and 6 using the SMART, PROST+HM, PROST, L+S, LLR, and MORASA methods. The corresponding error images for the reference estimated from fully sampled data are also shown. The nRMSEs are shown at the bottom of the error images.}
\label{fig5}
\end{figure*}

\section{RESULTS}
\subsection{Tissue Clustering Evaluation Experiment}
 $\text T_{1\rho}$-weighted images at TSL = 80 ms of the numerical phantom with different SNRs are shown in Fig.2(a) to represent low, medium, and high SNRs. Fig.2(b) shows the rank curves of the numerical and real phantom, and in vivo datasets. The first row of Fig.2(b) is for the fully sampled dataset, while the second is for reconstructing the undersampled data. 
\subsubsection{Numerical phantom dataset}  The first column of Fig.\ref{fig2}(b) shows rank curves of the numerical phantom. Since the numerical phantom was simulated using the bi-exponential relaxation model, the rank of $\mathbf G_\text{Block}$ and the mean rank of the Hankel matrices should be 2. For the fully sampled dataset, the rank of $\mathbf G_\text{Block}$ was maintained at 2 for all SNRs, while the mean rank of the Hankel matrices was greater than 2.5. For the accelerated dataset, the rank of $\mathbf G_\text{Block}$ was still maintained at 2, while the mean rank of the Hankel matrices for the accelerated dataset increased to nearly 3.
\subsubsection{Real phantom dataset} The second column of Fig. \ref{fig2} (b) shows rank curves of the real phantom dataset. Since each vial of the phantom was considered to be composed of a single component, the rank values of $\mathbf G_\text{Block}$ and the Hankel matrices should be 1. For the fully sampled dataset, the rank of $\mathbf G_\text{Block}$ was maintained at 1 for all SNRs, while the mean rank of the Hankel matrices was greater than 1. For the accelerated dataset, the rank of $\mathbf G_\text{Block}$ increased due to the residual artifacts, especially at low SNR (SNR = 25). The mean rank of the Hankel matrices for the accelerated dataset was even higher than 2.

\subsubsection{In vivo dataset} Two different tissues (denoted tissue 1 and tissue 2) were selected to evaluate the performance of the tissue clustering method. Tissue 1 was selected from tissue groups considered composed of a single-component and tissue 2 was selected from tissue groups considered composed of multi-components, according to the rank value of $\mathbf G_\text{Block}$. The third and fourth columns of Fig. \ref{fig2} (b) show the rank curves of these two tissues, respectively.
For tissue 1, the rank of $\mathbf G_\text{Block}$ was maintained at 1, while the mean rank of the  the Hankel matrices was higher than 2 for both fully sampled and accelerated datasets. For tissue 2, the rank of $\mathbf G_\text{Block}$ was maintained at 2 when $\text{SNR}\geq 30$ and increased at low SNR = 25. The mean rank of Hankel matrices was higher than 2 for both fully sampled and accelerated datasets.

The above results imply that the tissue clustering method is robust to noise and residual artifacts, and can improve the low-rank properties of the Hankel structure matrix.

\subsection{Retrospective 2D Reconstruction}
\subsubsection{Low acceleration with 4-Fold and 6-Fold }
\begin{table}[!htbp]
\caption{COMPARISONS OF DIFFERENT METHODS FOR THE RETROSPECTIVE 2D RECONSTRUCTIONS WITH DIFFERENT ACCELERATION FACTORS (R). THE BEST RESULTS ARE IN BOLD.}
\begin{center}
\setlength{\tabcolsep}{0.58mm}{
\begin{tabular}{cc|cccccc}
\hline \hline  & Metrics & SMART & PROST+HM & PROST & L+S & LLR & MORASA \\
\hline \multirow{3}{*}{ 4$\times$}& HFEN & \textbf{0.1858} & 0.2232  & 0.2359 & 0.2453 & 0.2645 & 0.2017 \\
& SSIM & \textbf{0.9805}  & 0.9762  & 0.9761 & 0.9737 & 0.9731 & 0.9780 \\
& PSNR & \textbf{43.2262} & 41.5372  & 41.1570 & 40.4791 & 39.8966 & 42.6131\\
\hline \multirow{3}{*}{6$\times$}& HFEN & \textbf{0.2224} & 0.2671  & 0.2727 & 0.2897 & 0.3170 & 0.2553 \\
& SSIM & \textbf{0.9761}  & 0.9701  & 0.9704 & 0.9670 & 0.9641 & 0.9715 \\
& PSNR & \textbf{42.0636} & 40.0560  & 40.1768 & 39.6337 & 39.0524 & 41.0307\\
\hline \hline
\end{tabular}}
\label{table 1}
\end{center}
\end{table}

The $\text T_{1\rho}$-weighted images (at TSL = 1 ms) from one volunteer reconstructed using the SMART, PROST+HM, PROST, L+S, LLR, and MORASA methods are shown in Fig. \ref{fig4}. Difference images between the reconstructed and reference images are displayed under the reconstructions, and the nRMSE values are placed below the difference images. Table \ref{table 1} lists the average HFEN, SSIM, and PSNR values for all reconstructed $\text T_{1\rho}$-weighted images at different TSLs. The estimated $\text T_{1\rho}$ maps from the reconstructed and difference images are shown in Fig. \ref{fig5}.

Supplementary information Fig. S2 shows the reconstructed $\text T_{1\rho}$-weighted images (at TSL = 1 ms) and the magnified images of the ROI from another volunteer using the SMART, RROST + HM, PROST, L + S, LLR, and MORASA methods at R = 4 and 6, respectively. The SMART method can better preserve the image resolution and finer details than the RROST + HM, PROST, and MORASA methods. The L + S and LLR methods have an image enhancement effect in the reconstructed images, particularly in the sulcus area, which is indicated by the yellow arrow. Compared to the methods exploiting spatial low-rankness (PROST, L + S, and LLR methods), the reconstruction can be improved by using parametric low-rankness based on high correlations in signal evolution in the reconstruction model.

\subsubsection{Higher acceleration with 10.2-Fold and 11.7-Fold } 
\begin{figure*}[!htbp]
\centering{\includegraphics[width=2\columnwidth]{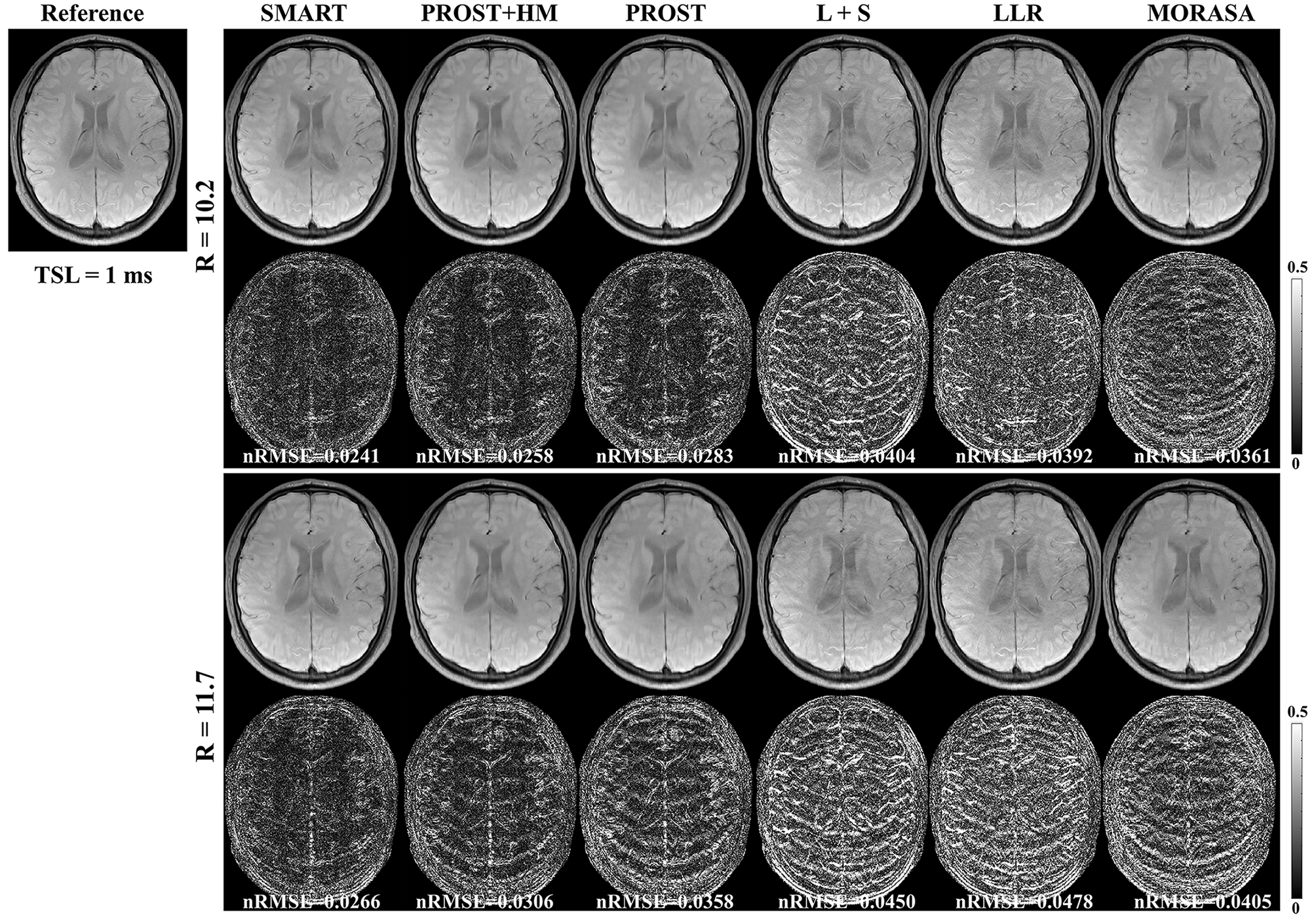}}
\caption{Reconstructed $\text T_{1\rho}$-weighted images from one 2D brain dataset at TSL = 1 ms with acceleration factors (R) = 10.2 and 11.7 using the SMART, PROST + HM, PROST, L + S, LLR, and MORASA methods. The corresponding error images for the reference image reconstructed from the fully sampled data are also shown. The error images are amplified by ten for visualization.  The nRMSEs are shown at the bottom of the error images.}
\label{2D higher acceleration result}
\end{figure*}

The reconstructed $\text T_{1\rho}$-weighted images and the corresponding error images with R = 10.2 and 11.7 are shown in Fig. \ref{2D higher acceleration result}. The nRMSEs are shown at the bottom of each error image. The reconstructed  $\text T_{1\rho}$-weighted images using the SMART method have no notable artifacts, even at a high acceleration factor of up to 11.7. The related errors remain noise-like, and the nRMSEs are less than 3$\%$. Noticeable aliasing artifacts can be observed from the images or error maps using the PROST+HM, PROST, L+S, LLR, and MORASA methods at different TSLs. The corresponding $\text T_{1\rho}$ maps are shown in supplementary information Fig. S3.
Table \ref{table 2} shows the average HFEN, SSIM, and PSNR values for all reconstructed $\text T_{1\rho}$-weighted images. The proposed SMART method qualitatively achieves the best performance among the six methods, especially at R = 11.7.

\begin{table}[htbp]
\caption{COMPARISONS OF DIFFERENT METHODS FOR THE RETROSPECTIVE 2D RECONSTRUCTIONS WITH DIFFERENT ACCELERATION FACTORS (R). THE BEST RESULTS ARE IN BOLD.}
\begin{center}
\setlength{\tabcolsep}{0.6mm}{
\begin{tabular}{cc|cccccc}
\hline \hline  & Metrics  & SMART & PROST+HM & PROST & L+S & LLR & MORASA \\
\hline \multirow{3}{*}{ 10.2$\times$}& HFEN & \textbf{0.2558} & 0.2953  & 0.3360 & 0.4057 & 0.4369 & 0.3459 \\
& SSIM & \textbf{0.9728}  & 0.9688  & 0.9637 & 0.9582 & 0.9542 & 0.9674 \\
& PSNR & \textbf{40.3712} & 39.4523  & 38.5946 & 36.8233 & 35.3039 & 37.8079\\
\hline 
\multirow{3}{*}{11.7$\times$}& HFEN & 
\textbf{0.2817} & 0.3460  & 0.4134 & 0.4527 & 0.5148 & 0.3762 \\
& SSIM & \textbf{0.9701}  & 0.9640  & 0.9578 & 0.9533 & 0.9472 & 0.9634 \\
& PSNR & \textbf{39.6822} & 38.1819  & 36.8288 & 35.9089 & 33.6896 & 37.2335\\
\hline \hline
\end{tabular}}
\label{table 2}
\end{center}
\end{table}

\begin{figure*}[!htbp]
\centering{\includegraphics[width=2\columnwidth]{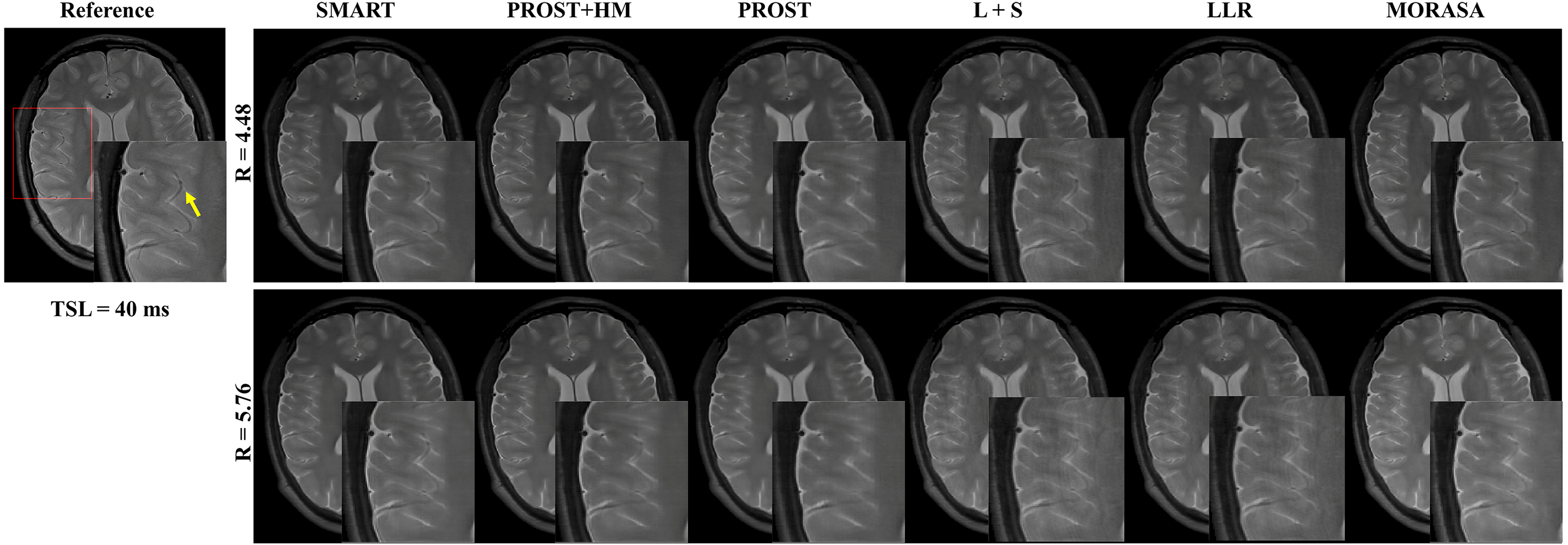}}
\caption{Reconstructed $\text T_{1\rho}$-weighted images and magnified images from prospective 2D brain dataset at TSL = 40 ms with R = 4.48 and 5.76 using the SMART, PROST + HM, PROST, L + S, LLR, and MORASA methods. Aliasing artifacts can be seen in magnified images of reconstructions using L+S and LLR methods, and the yellow arrow shows a blood vessel in the reference image.}
\label{prospective result}
\end{figure*}

\begin{figure*}[!htbp]
\centering{\includegraphics[width=2\columnwidth]{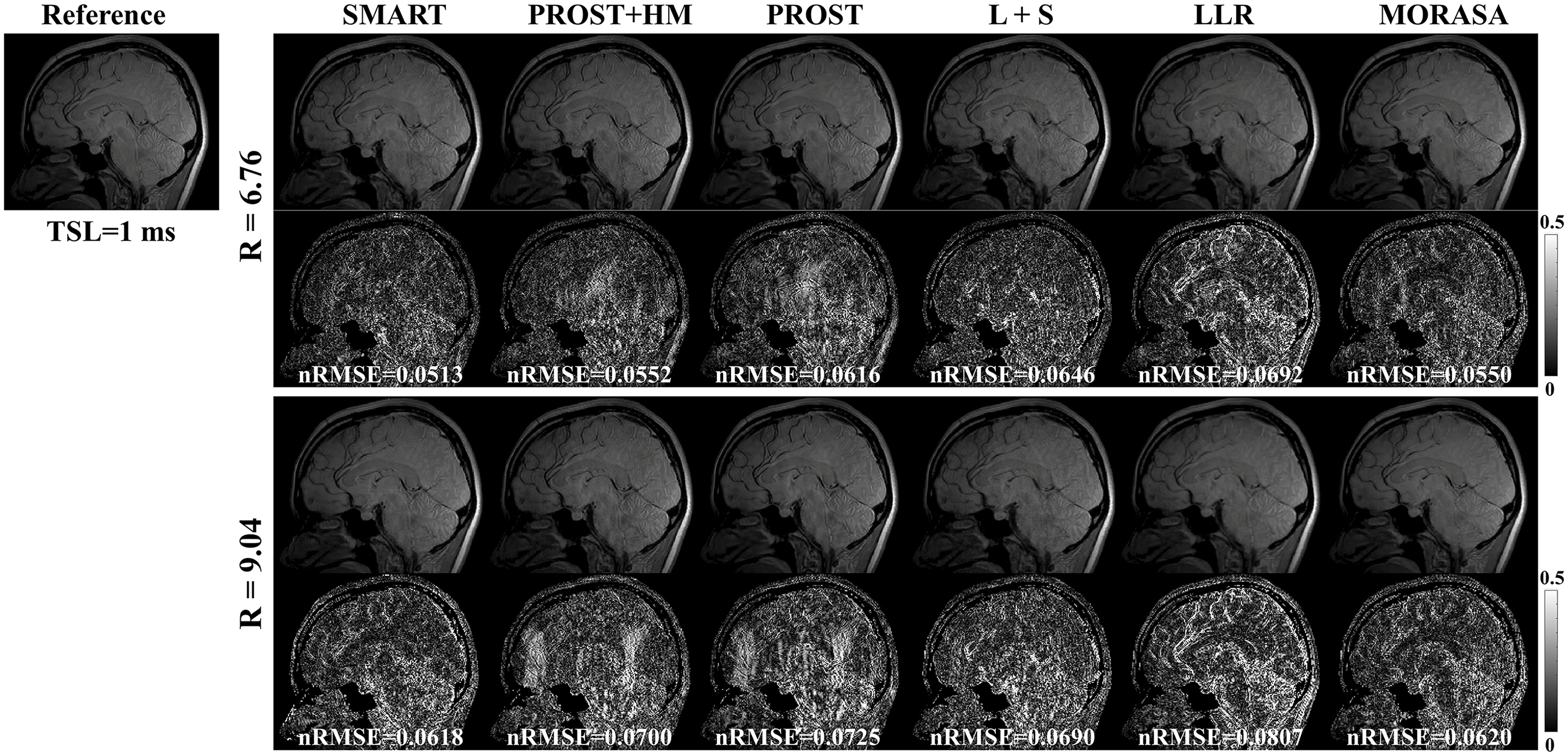}}
\caption{Reconstructed $\text T_{1\rho}$-weighted images from one slice of the 3D brain dataset at TSL = 1 ms with acceleration factors (R) = 6.76 and 9.04 using the SMART, PROST + HM, PROST, L + S, LLR, and MORASA methods. The corresponding error images for the reference estimated from fully sampled data are also shown. The error images are amplified by ten for visualization. The nRMSEs are shown at the bottom of the error images.}
\label{fig7}
\end{figure*}

\subsection{Prospective 2D Reconstruction}
Fig. \ref{prospective result} shows the prospective reconstructed $\text T_{1\rho}$-weighted images (at TSL = 40 ms) from another volunteer and the magnified images using the SMART, PROST + HM, PROST, L + S, LLR, and MORASA methods. Visual artifacts can be observed in the magnified images of reconstructions at all accelerating factors using the L + S and LLR methods. The images reconstructed using the PROST + HM and PROST methods were blurred compared to those reconstructed using the SMART method. Some image details (namely, the blood vessel marked in a yellow arrow) can narrowly be seen in the reconstructions using the PROST + HM and MORASA methods and disappeared in the reconstruction using the PROST method at R = 5.76. In contrast, the image details were well preserved using the SMART method. The $\text T_{1\rho}$ maps reconstructed using the methods above are shown in supplementary information Fig. S4. Similar conclusions can be drawn from the $\text T_{1\rho}$ maps.

\subsection{3D Reconstruction}
\subsubsection{Low acceleration with 6.76-Fold and 9.04-Fold }
The $\text T_{1\rho}$-weighted images (at TSL = 1 ms) from one slice of the reconstructed 3D images using the SMART, PROST + HM PROST, L + S, LLR, and MORASA methods for R = 6.76 and 9.04 are shown in Fig.\ref{fig7}. The $\text T_{1\rho} $ maps reconstructed using the methods above are shown in supplementary information Fig. S5. 
\subsubsection{Low acceleration with 11.26-Fold and 13.21-Fold}
\begin{figure*}
\centering{\includegraphics[width=2\columnwidth]{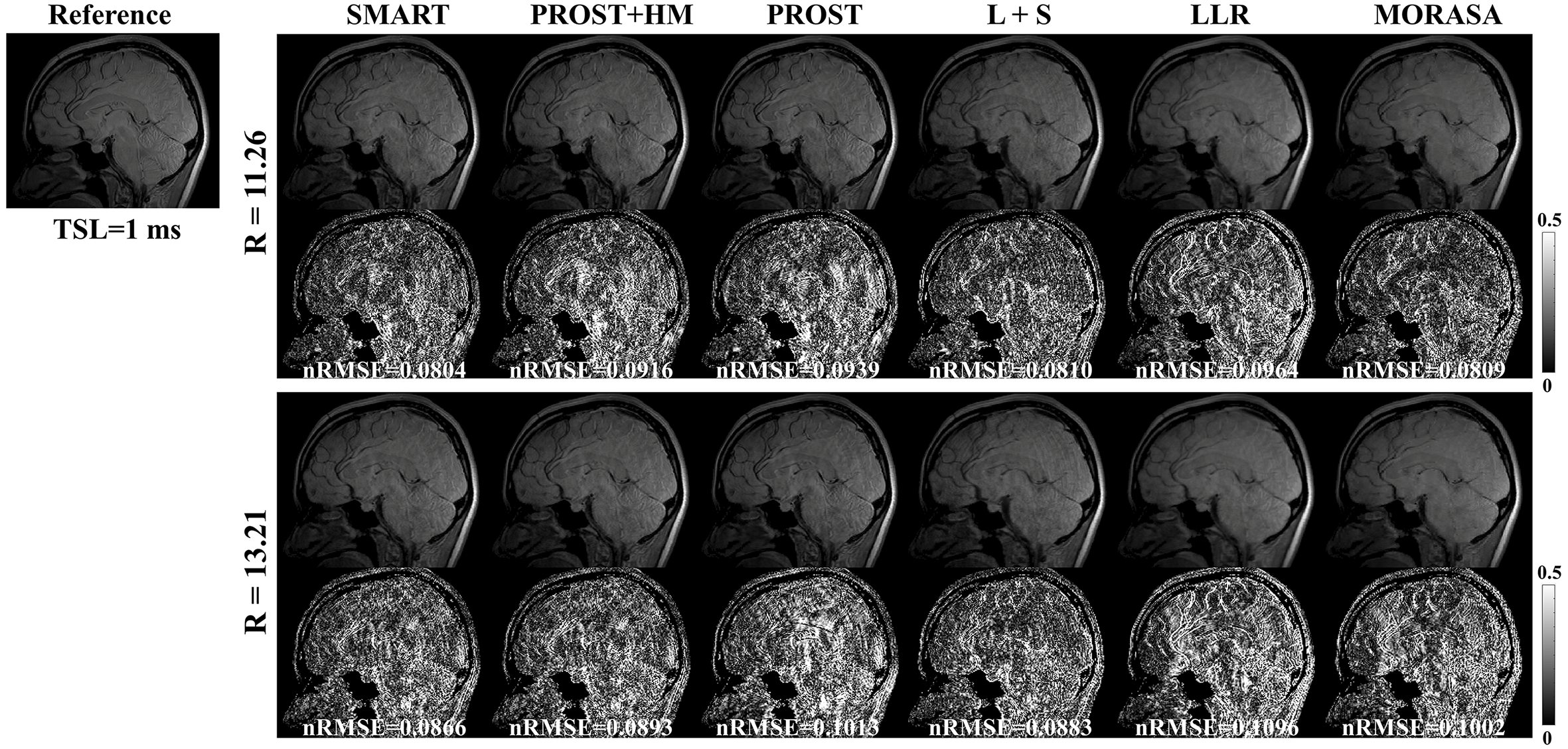}}
\caption{ Reconstructed $\text T_{1\rho}$-weighted images from one slice of the 3D brain dataset at TSL = 1 ms with acceleration factors (R) =11.26 and 13.21 using the SMART, PROST + HM, PROST, L + S, LLR, and MORASA methods. The corresponding error images for the reference estimated from fully sampled data are also shown. The error images are amplified by ten for visualization. The nRMSEs are shown at the bottom of the error images. }
\label{3D higher acceleration result}
\end{figure*}

Fig. \ref{3D higher acceleration result} shows the reconstructed $\text T_{1\rho}$-weighted images and the related error maps at TSL = 1 ms with R = 11.26 and 13.21 using the six methods in 3D imaging. The estimated $\text T_{1\rho}$ maps from the reconstructions are shown in supplementary information Fig. S6. At high acceleration, the low-rank tensor-based reconstruction methods (SMART, PROST, and  PROST+HM) show better detail retention than the low-rank matrix-based methods (LLR and MORASA). The SMART method still achieves lower nRMSE than other methods at higher accelerations.
 \subsection{Parameter Selection }
\begin{figure}[!htbp]
\centering{\includegraphics[width=\columnwidth]{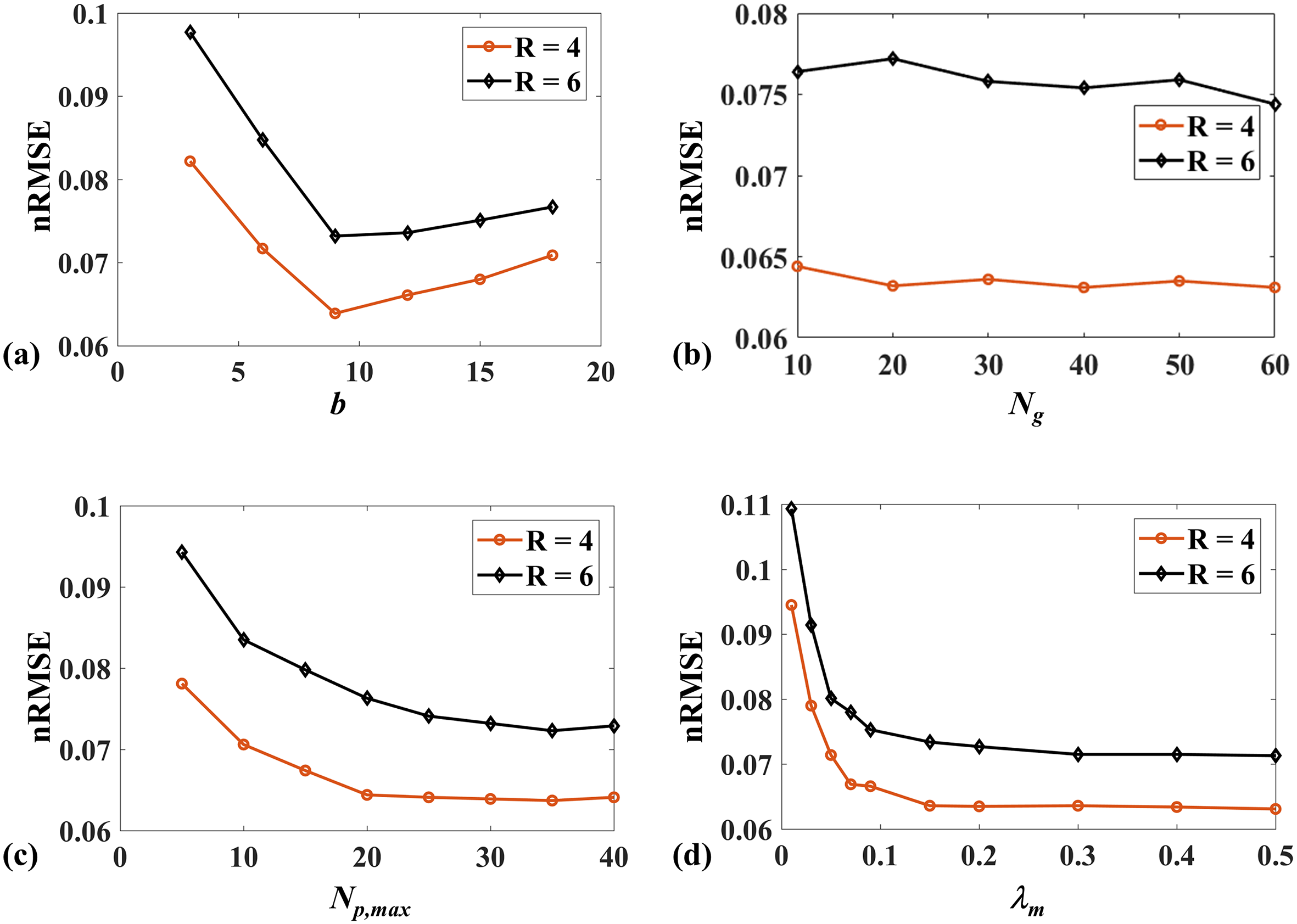}}
\caption{nRMSE curves for the reconstructions of SMART using different parameters related to the spatial and parametric tensors.}
\label{parameterselection}
\end{figure}

Fig.\ref{parameterselection} shows the effects of four parameters of the spatial and parametric tensor construction in 2D imaging.
As shown in Fig. \ref{parameterselection}(a), the smallest nRMSE is achieved at $b = 9$. When the patch size decreases from 9 to 3, the nRMSE increases rapidly due to the small image patches’ limited spatial modeling ability. When the patch size is larger than 9, the reconstruction performance slowly degrades since the discrepancy among the patches increases, causing the reduced low rankness of the constructed spatial tensor. The nRMSE of reconstruction decreases with the increase of $ N_{p,max}$, as shown in Fig. \ref{parameterselection}(c), but the trend slows down when $ N_{p,max}\geq 30$. Considering the computational complexity, we chose $ N_{p,max}=30$ in this study. Fig. \ref{parameterselection}(d) shows that $\lambda_m$ needs to be larger than 0.15; otherwise, the nRMSE will significantly increase. As shown in Fig. \ref{parameterselection}(b), changing $ N_{g}$ has little effect on the nRMSE of the reconstruction for R = 4. In contrast, large $ N_{g}$ shows improved performance for the higher acceleration factor of 6. Therefore, $ N_{g}=60$ was used in this study.

Supplementary information Fig. S7 shows the representative zero-filling reconstruction with R = 6, the intermediate reconstruction at the 12th iteration, the parameter histogram, the estimated $\text T_{1\rho}$ map,  the corresponding tissue groups, and the ranks of tissue groups. The initial $\text T_{1\rho}$-weighted image of the zero-filling reconstruction shows strong aliasing artifact, and the $\text T_{1\rho}$ map estimated from zero-filling images is quite blurred. The quality of $\text T_{1\rho}$-weighted image and $\text T_{1\rho}$ map was significantly improved at the 12th iteration. After iteration, the distribution of the parameter histogram also changed (shown in Fig. S7(b)). In the figure of tissue groups, the pixel intensities along the pixel number direction are not as uniform as those at the 12th iteration. The ranks in several groups are reduced from 2 to 1.
\subsection{Ablation Experiment }
\begin{figure}[!htbp]
\centering{\includegraphics[width=\columnwidth]{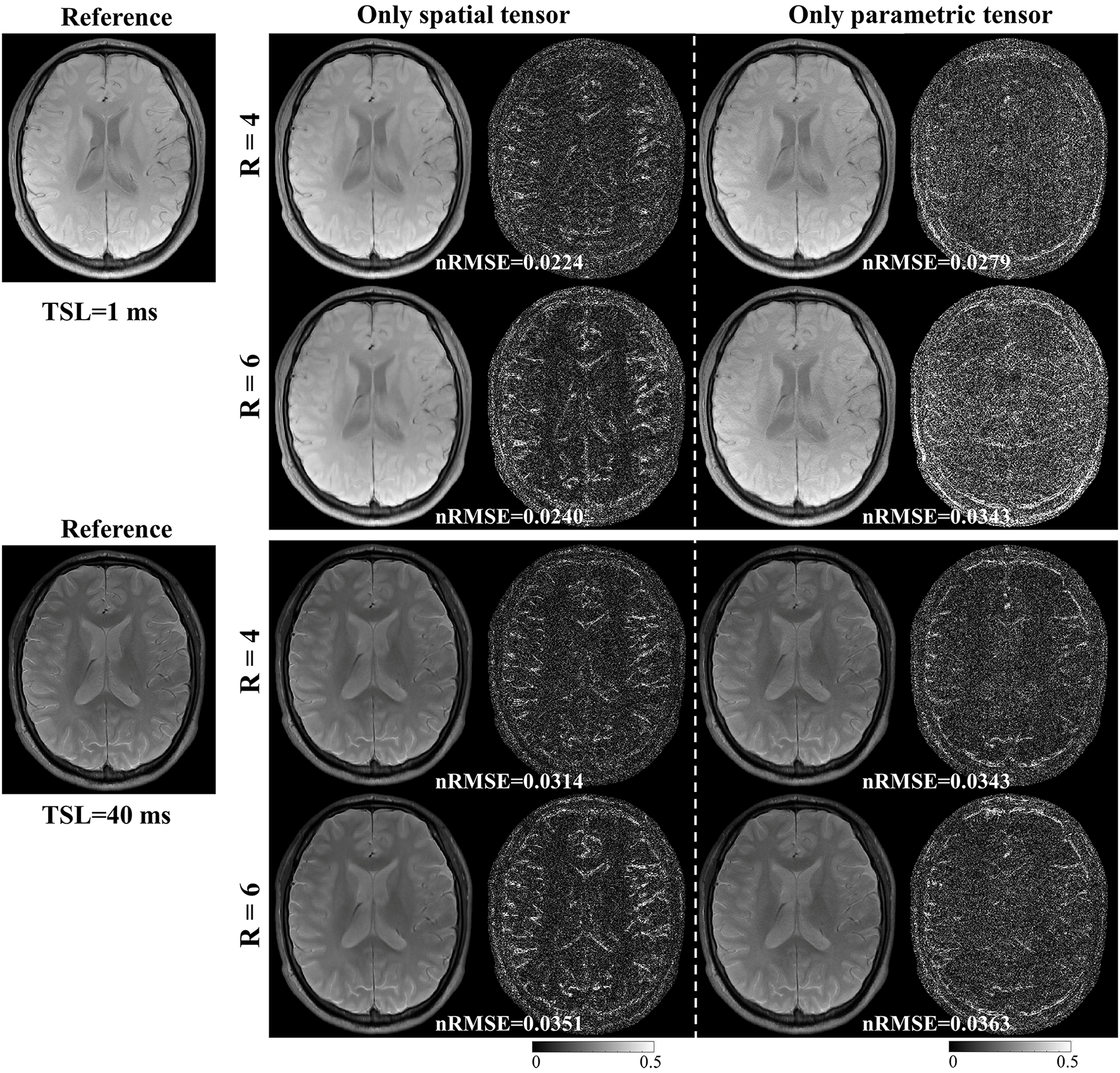}}
\caption{The reconstructed images at TSL = 1 ms and 40 ms for the reconstruction models with only spatial tensor and only parametric tensor at R = 4 and 6, respectively. Each error image is placed on the right of each reconstructed image along with the nRMSE. The error images are amplified by ten for visualization.}
\label{ablation result}
\end{figure}

Fig. \ref{ablation result} shows the reconstructed images for models 1 and 2 with R = 4 and 6, respectively. For the results of model 1, the aliasing artifacts due to undersampling can be well removed from the reconstructed images by applying the spatial tensor, but the reconstructed images are blurry. In contrast, the sharpness of images reconstructed by model 2 is improved, but aliasing artifacts still exist. These results are in line with our hypothesis that the parametric group-based  tensor can help improve image sharpness and preserve more image details in the reconstructions.
\section{DISCUSSION}
\par This study proposed the SMART method using simultaneous spatial patch-based and parametric group-based low-rank tensors for accelerated MR $\text T_{1\rho}$ mapping. We demonstrated that the proposed method could recover high-quality images from highly undersampled data in 2D and 3D imaging. The high local and nonlocal redundancies and similarities between the contrast images were exploited using the spatial patch-based low-rank tensor through HOSVD. Compared with transforms that use fixed bases, such as discrete cosine and wavelet, the HOSVD bases $\textbf U^{(n)}$ are learned from the multidimensional data and thus more adaptive to the structures in the data. This adaptive nature enables HOSVD to be an excellent sparsifying transform for image reconstruction\cite{RN644}. Similar overlapping patches are used to form the spatial low-rank tensors as in the PROST method. The optimization subproblem for the tensors can be addressed in the high-order denoising process. The denoised image is obtained by averaging different estimates, which may lead to blurring in the reconstructions. In the proposed SMART method, in addition to the spatial tensor, we jointly used a parametric group-based low-rank tensor in the reconstruction model, which exploits the low rank in spatial and temporal directions. The parametric tensor is formed non-overlapping, which exhibits stronger low-rankness based on high correlations in tissues of similar signal evolution. It may help improve image sharpness and preserve more image details in the reconstructions. Therefore, the SMART method can perform better than existing spatial patch-based methods.

\par Meanwhile, the SMART method exhibited high robustness in prospective experiments and can be extended to other quantitative imaging methods, such as $\text T_2$ mapping and $\text T_1$ mapping. 

\subsection{Parameters Selection for the SMART Method}
\begin{figure}[!htbp]
\centering{\includegraphics[width=\columnwidth]{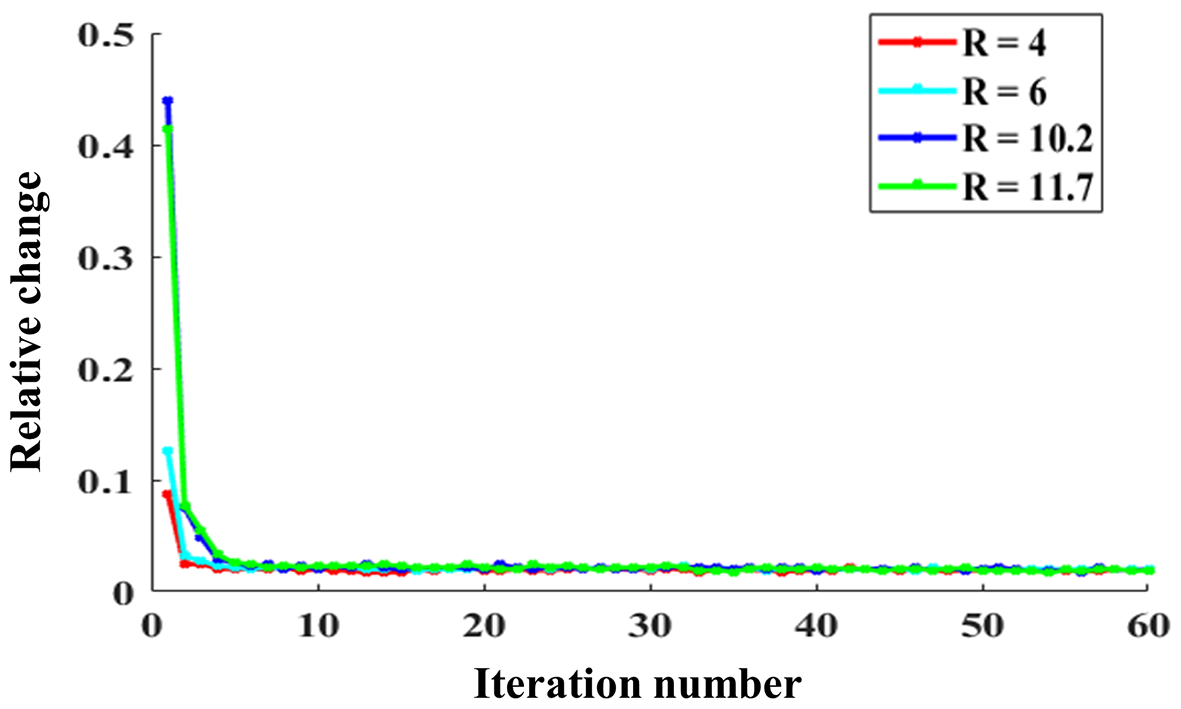}}
\caption{Relative change in the SMART reconstruction between the two consecutive ADMM iterations for one of the 2D brain datasets with R = 4, 6, 10.2, and 11.7.}
\label{relativechange}
\end{figure}

In this study, all the reconstruction parameters included in the ADMM and CG algorithms were selected empirically. The relative change in the solution between two consecutive iterations was used to show the convergence numerically and is defined as follows: 
\begin{equation}
\left \|\mathbf X_\textit{iter}-\mathbf X_ \textit{iter-1 }\right \|_F / \left \|\mathbf X_ \textit{iter-1 }\right \|_F
\end{equation}
where $\mathbf X_\textit{iter}$ denotes the reconstructed image at the \textit{iter}th ADMM iteration. Fig. \ref{relativechange} shows the relative changes of 2D imaging with R = 4, 6, 10.2, and 11.7. The relative change rapidly stabilizes within a few iteration numbers and has no significant alterations when the iteration number reaches 15. Additionally, we changed the CG iteration number from 7 to 20 to test its effects on the reconstruction. The nRMSE variations of the reconstructed images with different CG iteration numbers were less than 1\%, indicating the CG iteration number may have little effect on reconstruction after several iterations.

 We used several tricks to speed up the reconstruction in our experiments. For the patch extraction step, a patch offset with three was used to reduce the number of searched patches. Also, the patch extraction step was implemented in a graphics processing unit (GPU). The time consumption of this step was reduced from 9.5s to 0.14s for the 2D patch extraction and from 600 mins to 171s for the 3D patch extraction. The memory footprints were 411 MB and 821 MB for the 2D and 3D patch extractions. The $\text T_{1\rho}$ estimation for the histogram analysis of the tissue clustering was applied every three iterations to speed up the reconstruction. Additionally, parallel computing was applied in the HOSVD denoising step of tensors $\mathcal{T}$ and $\mathcal{Z}$ for each spatial and parametric tensor. The reconstruction times of the six methods are listed in supplementary information Table S3.

\subsection{2D Patch vs. 3D Patch in the 3D Imaging Application}
\begin{figure}[!htbp]
\centering{\includegraphics[width=\columnwidth]{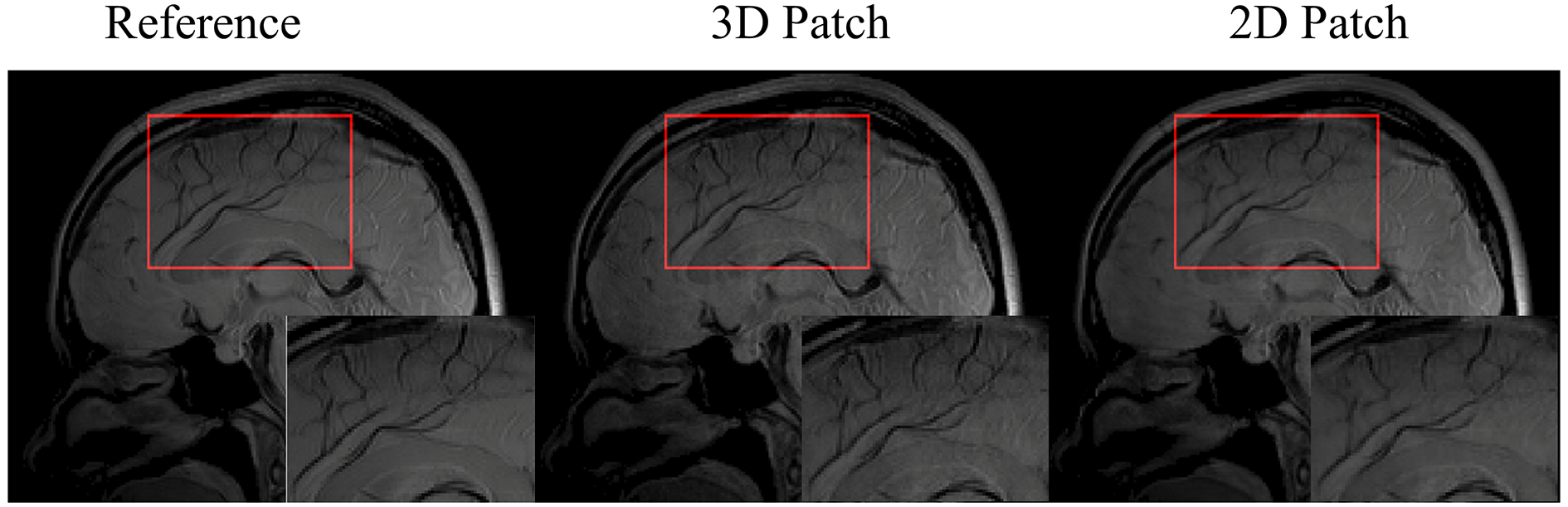}}
\caption{ One slice of the reconstructed 3D images at TSL = 1 ms using 2D and 3D patch extraction, respectively, with R = 9.04. }
\label{fig12}
\end{figure}

For 3D imaging, the images can be reconstructed slice-by-slice using the SMART method, where the 2D patch extraction is implemented. Compared to the 2D patch extraction, the correlation information between the slices can be utilized to improve the reconstruction. Therefore, in this study, the 3D patch extraction was selected in the SMART method. Fig. \ref{fig12} shows a slice of the reconstructed images at R = 9.04 using the 2D and 3D patch extraction, respectively. The reference images from the fully sampled data and magnified images are also shown. As shown in the magnified images, the 3D patch extraction displays better image resolution and detail, particularly in the regions indicated by the red frames. 
\subsection{Comparison with Previous Studies}
Previous studies \cite{RN2011,RN2012} have utilized convolutional sparse coding (CSC) to reconstruct images from undersampled data. Thanh et al.\cite{RN2011} proposed a filter-based dictionary learning method using a 3D CSC to recover the high-frequency information of the MRI images. The CSC employed a sparse representation of the entire image computed by the sum of a set of convolutions with dictionary filters instead of dividing an image into overlapped patches. It can solve high computational complexity and long-running time problems due to patch-based approaches and avoid artifacts caused by patch aggregation. However, filters must be carefully chosen, and artifacts may appear in some reconstructed images due to inaccurate filters\cite{RN2012}.

Low rankness has been widely used in fast MR parametric mapping. The annihilating filter-based low-rank Hankel matrix method (ALOHA)\cite{RN594} exploits the transform domain sparsity and the coil sensitivity smoothness, representing a low-rank Hankel structured matrix in the weighted $k$-space domain. The prior information used in ALOHA differs from the image domain redundancy utilized in our study. Specifically, the spatial and temporal redundancies are mainly represented as a summation of exponentials and the Hankel matrix at each voxel with a low-rank property. Multi-scale low-rank with different patch sizes was also used to capture global and local information. It has a more compact signal representation and may further improve MR reconstruction. Frank Ong et al.\cite{RN2013, RN2014} have applied multi-scale low-rank in face recognition preprocessing and dynamic contrast-enhanced MRI. However, when constructing multiple low-rank tensors, the memory footprint is large, and time consumption is increased several times compared with a single patch size. Another problem is that the multi-scale low-rank decomposition is not translation invariant; shifting the input changes the resulting decomposition. This translation variant nature often creates blocking artifacts near the block boundaries, which can be visually jarring. Introducing overlapping partitions of the patches so that the overall algorithm is translation invariant can remove these artifacts. The proposed SMART method uses a single patch size with overlapped patches considering the limits of memory and reconstruction time. 

\subsection{Limitations of SMART}
In this study, the setup of the proposed method is non-convex, and the patch selection operator is time-varying. The convergence of the whole algorithm cannot be completely guaranteed. However, based on our experimental results, the proposed method works well despite the lack of a convergence guarantee. We found that after several iterations, the relative change of reconstructions between two consecutive iterations and the patch extract selection varies slightly. Generally, this estimation based on the patch structure remains consistent along with the increasing iteration. In addition, the block matching algorithm for patch extraction is heuristic, and the current selection strategy may not present the best clusters. In future work, patches obtained through a sliding operation throughout the data may be used to construct the high order tensor. The local and nonlocal redundancies and similarities between the contrast images can be exploited using the tensor dictionary learning method by jointly applying the low-rank constraints on the dictionaries, and the sparse constraints on the core coefficient tensors \cite{RN2006}. This may further improves the reconstruction performance.

 The iterative method’s long reconstruction time significantly hinders the proposed method’s clinical application. However, there are two potential solutions to this problem. One solution is using parallel computing via GPU. We have already implemented the patch extraction in GPU. 
Although the reconstruction time is still too long for clinical use, we plan to improve the speed of the SMART method by leveraging parallel computing technology, such as multi-GPU and cluster systems. The other solution is the deep learning-based reconstruction method. In recent years, the unrolling-based strategy has established a bridge from traditional iterative CS reconstruction to deep neural network design \cite{RN2007}. Previous studies have shown that an unrolling-based deep network significantly reduced the reconstruction time and improved the reconstruction quality through the learnable regularizations and parameters. However, deep learning-based reconstruction requires a considerable amount of high-quality training data, and the network performance may degrade due to the motion artifact during the data acquisition. Training data collection is critical and will be the bottleneck of the deep learning-based reconstruction method. In our future work, we will try both strategies to promote the clinical use of the proposed method.
\section{CONCLUSIONS}
This study explores low-rankness through high-order tensor in MR $\text T_{1\rho}$ mapping to obtain improved reconstruction results. In particular, we propose a method that simultaneously uses spatial patch-based and parametric group-based low-rank tensors to reconstruct images from highly undersampled data. The spatial patch-based low-rank tensor exploits the high local and nonlocal redundancies and the similarities between the contrast images. The parametric group-based low-rank tensor, which integrates similar exponential behavior of the image signals, is jointly used to enforce the multidimensional low-rankness in the reconstruction process. Experimental results in both 2D and 3D imaging scenarios show that the proposed method can improve the reconstruction results qualitatively and quantitatively.

\bibliographystyle{IEEEtran}
\bibliography{tmi}

\section*{Supplementary Information }













\setcounter{table}{0}
\setcounter{figure}{0}
\renewcommand\thefigure{S\arabic{figure}}
\renewcommand\thetable{S\arabic{table}} 

\subsection{Data Acquisition Information}The 2D dataset was collected on a 3T scanner (TIM TRIO,Siemens, Erlangen, Germany) using a 12-channel head coil from three volunteers (age: 24 ± 2 years). The 3D dataset was collected on a 3T scanner (uMR 790, United Imaging Healthcare, Shanghai, China) using a 32-channel head coil from two healthy volunteers (age: 26 ± 1 years). $\text T_{1\rho}$-weighted images were acquired using the fast spin echo (FSE) sequence for 2D imaging and the modulated flip angle technique in refocused imaging with an extended echo train (MATRIX) sequence for 3D imaging, both using a spin-lock pulse at the beginning of the sequence to generate the $\text T_{1\rho}$-weighting \cite{RN680,RN687}. The main imaging parameters that used in the 2D and 3D imaging applications are listed in Supplementary information Table \ref{table S1}. 
\par The 2D phantom dataset was also collected a 3T scanner (TIM TRIO,Siemens, Erlangen, Germany) using a 12-channel head coil. The imaging sequence was the same as used in the in vivo data acquisition. To obtain an accurate $\text T_{1\rho}$ estimation of the phantom, a longer TR of 4000 ms was used. Therefore, the scan time of the phantom data acquisition was longer than that of the in vivo data acquisition. 
\subsection{Reconstruction Parameters } The reconstruction parameters that used the SMART method are listed in Supplementary information Table \ref{table S2}. In the L + S method, the singular-value thresholding and soft-thresholding algorithms were used to solve for the L and S components, respectively. The ratio for singular-value thresholding was set as 0.02, whereas that for soft-thresholding was set as [0.02, 0.025, 0.025, 0.035, and 0.035] to achieve optimal performance. In the LLR method, the iteration numbers were set as 40, 50, and 60 for R = 4 and 6, respectively. The block size was set as $8 \times 8$, and the ratio of the largest singular value used for the threshold of the singular value decomposition was initially set as 0.03, reduced to 0.01 after 10 iterations, and finally reduced to 0.005 in the final 10 iterations. The parameters for PROST and PROST + HM were the same as those for SMART. In the MORASA, the rank of the Casorati matrix was 2. The threshold for the wavelet coefficients and that of the Hankel matrix were 0.01 and 0.03, respectively.

\subsection{Real Phantom} 
\begin{figure*}[!htbp]
\centering{\includegraphics[width=1.8\columnwidth]{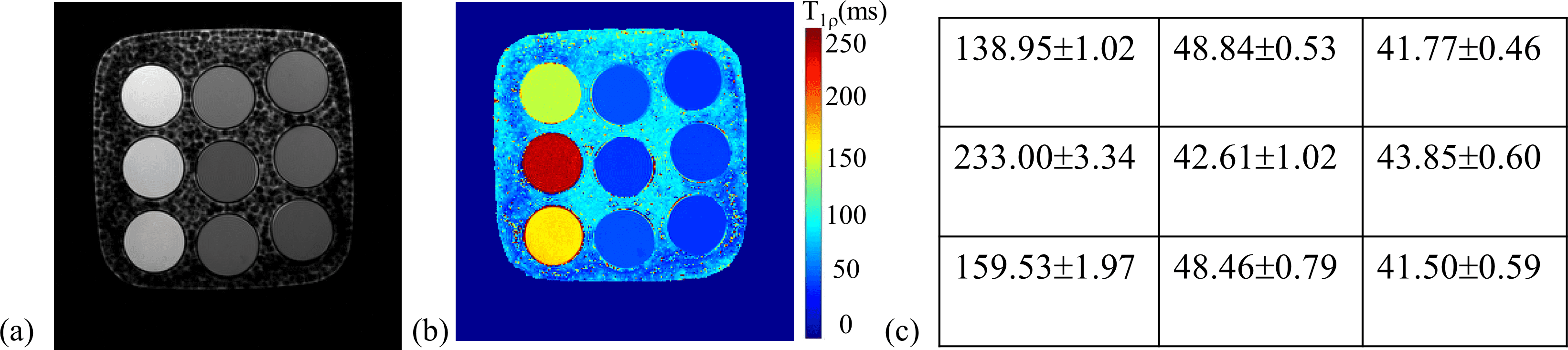}}
\caption{ The $\text T_{1\rho}$-weighted  image of the real phantom at TSL = 40 ms (a), the obtained $\text T_{1\rho}$ map (b), and the estimated $\text T_{1\rho}$ values (unit: ms) of the nine vials (c).}
\label{real_phantom_figure}
\end{figure*}

Supplementary Information Fig.\ref{real_phantom_figure}(a-c)   shows the  $\text T_{1\rho}$-weighted image  of the real phantom at TSL = 40 ms, the obtained $\text T_{1\rho}$ map, and the estimated $\text T_{1\rho}$ values of the nine vials, respectively. 

\subsection{Retrospective 2D Reconstruction}
\subsubsection{Low acceleration}
\begin{figure*}[!htbp]
\centering{\includegraphics[width=2\columnwidth]{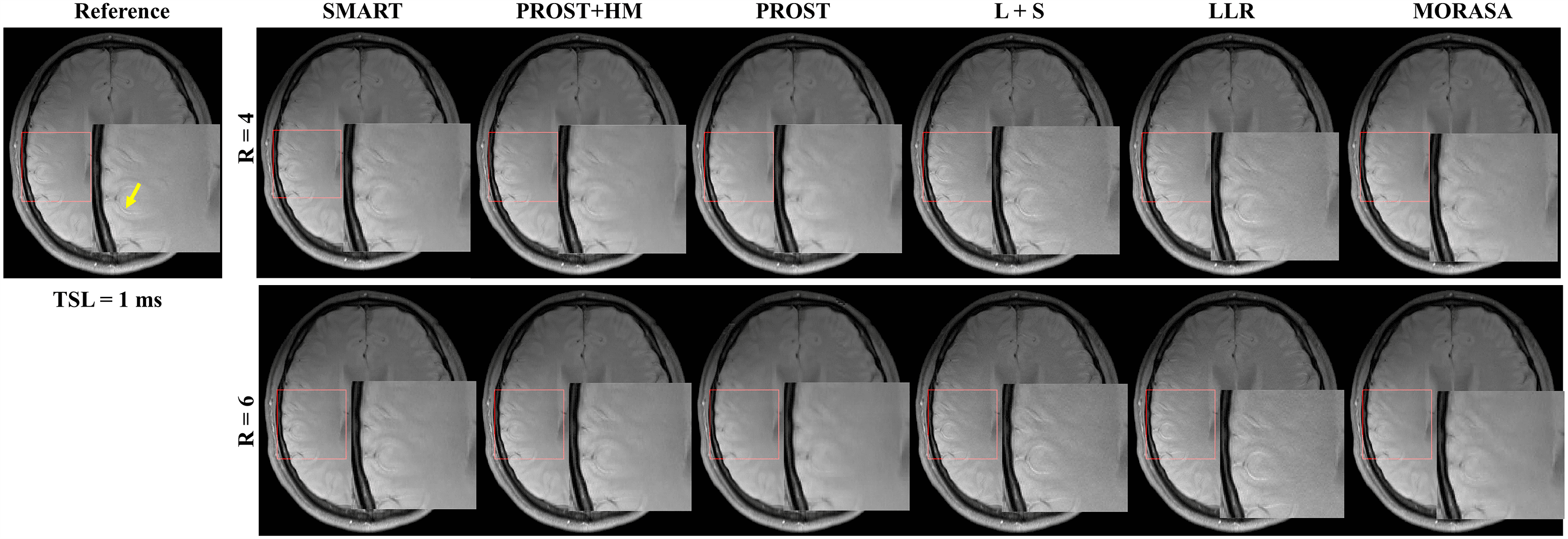}}
\caption{ Reconstructed $\text T_{1\rho}$-weighted images from another 2D brain dataset at TSL = 1 ms with acceleration factors (R) = 4 and 6 using the SMART, PROST + HM, PROST, L + S, LLR, and MORASA methods. The magnified figures of region of interest (denoted by the red box) are shown at the right bottom of each reconstructed image. The yellow arrow shows sulcus area in the reference image.}
\label{figure S1}
\end{figure*}

Supplementary  information Fig. \ref{figure S1} shows the reconstructed $\text T_{1\rho}$-weighted images (at TSL = 1 ms) and the magnified images of the region of interest from another volunteer using the SMART, RROST + HM, PROST, L + S, LLR, and MORASA methods at R = 4 and 6, respectively. The SMART method can better preserve the image resolution and finer details than the RROST + HM, PROST, and MORASA methods. The L + S and LLR methods have an image enhancement effect in the reconstructed images, particularly in the sulcus area, which are indicated by the red arrows. Compared to the methods exploiting spatial low-rankness (that is, the PROST, L + S, and LLR methods), the reconstruction can be improved by using parametric low-rankness based on high correlations in signal evolution in the reconstruction model.

\subsubsection{Higher acceleration}
Supplementary  information Fig.\ref{figure S4} compares the $\text T_{1\rho}$ maps and the corresponding error maps obtained using different methods for R = 10.2 and 11.7. It can be seen that SMART generates the closest $\text T_{1\rho}$ map to the reference with the lowest nRMSE.The $\text T_{1\rho}$ maps obtained using L+S and LLR show obvious blurring artifacts, and $\text T_{1\rho}$ maps obtained using MORASA show aliasing artifacts (indicated with red arrows). The Patch-based low-rank tensor spatial tensor can reduce aliasing artifacts and preserve more image details. SMART further improves the reconstruction performance and $\text T_{1\rho}$ estimation accuracy by using the parametric tensor compared with PROST+HM and PROST. 

\subsection{Prospective 2D Reconstruction}
Supplementary information Fig. \ref{figure S2} shows the corresponding $\text T_{1\rho}$ maps reconstructed using the SMART, PROST + HM, PROST, L + S, LLR, and MORASA methods and the error maps between the reconstructed $\text T_{1\rho}$ map and the reference map. The $\text T_{1\rho}$ maps reconstructed using the SMART method exhibited superior reconstruction performance compared to those reconstructed using the other four methods, with the smallest nRMSEs and error maps for both acceleration factors.

\subsection{Retrospective 3D Reconstruction}
\subsubsection{low acceleration}The $\text T_{1\rho}$ maps reconstructed using the SMART, PROST + HM, PROST, L + S, LLR, and MORASA methods are shown in Supplementary information Fig. \ref{figure S3}.
\subsubsection{Higher acceleration}
\begin{figure*}[!htbp]
\centering{\includegraphics[width=1.8\columnwidth]{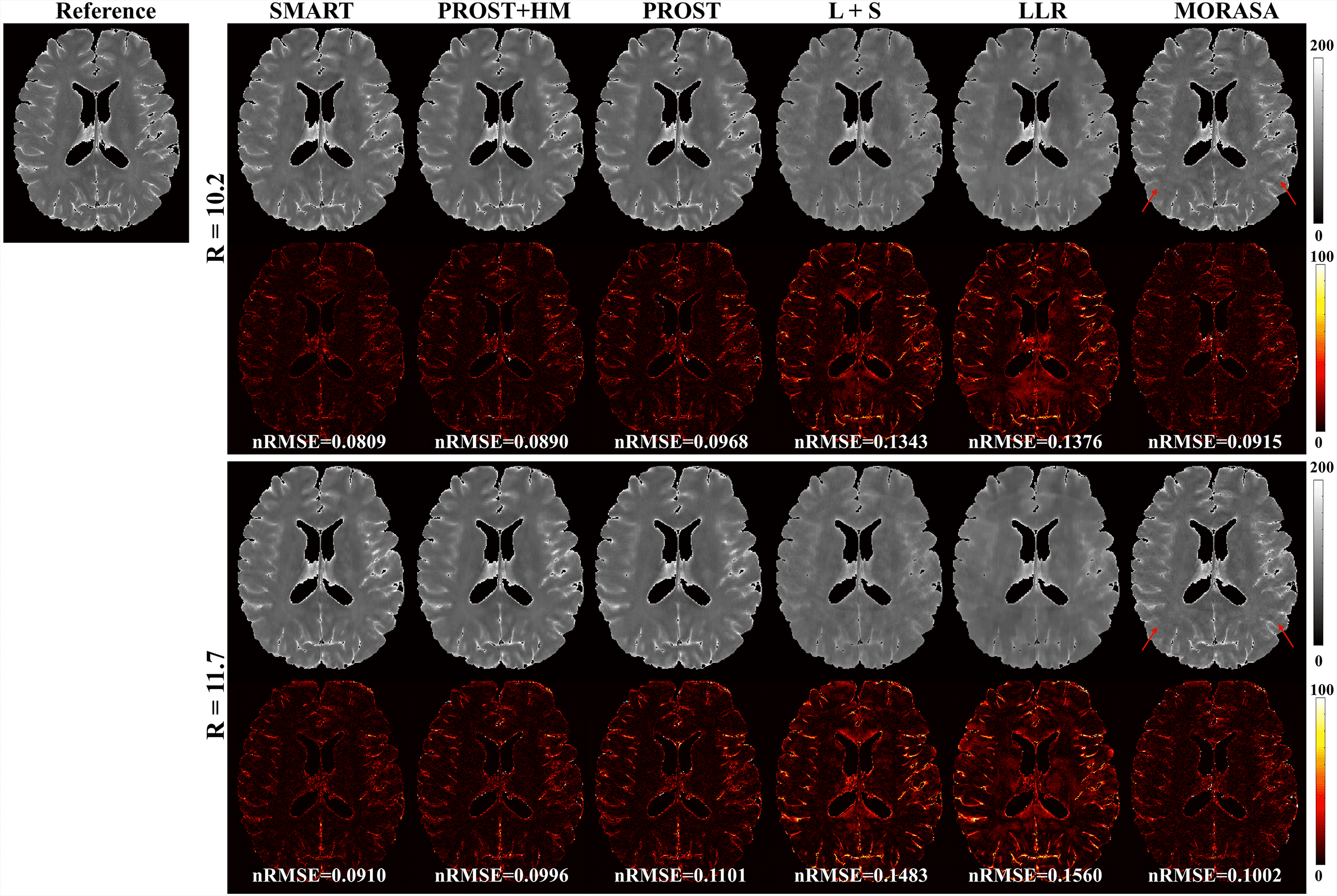}}
\caption{ $\text T_{1\rho}$ maps estimated from one 2D brain dataset with acceleration factors (R) = 10.2 and 11.7 using the SMART, PROST + HM, PROST, L + S, LLR, and MORASA methods. The corresponding error images with respect to the reference estimated from fully sampled data are also shown. The nRMSEs are shown at the bottom of the error images. Red arrows show the aliasing artifacts in reconstructions using the MORASA method.}
\label{figure S4}
\end{figure*}

\begin{figure*}[!htbp]
\centering{\includegraphics[width=2\columnwidth]{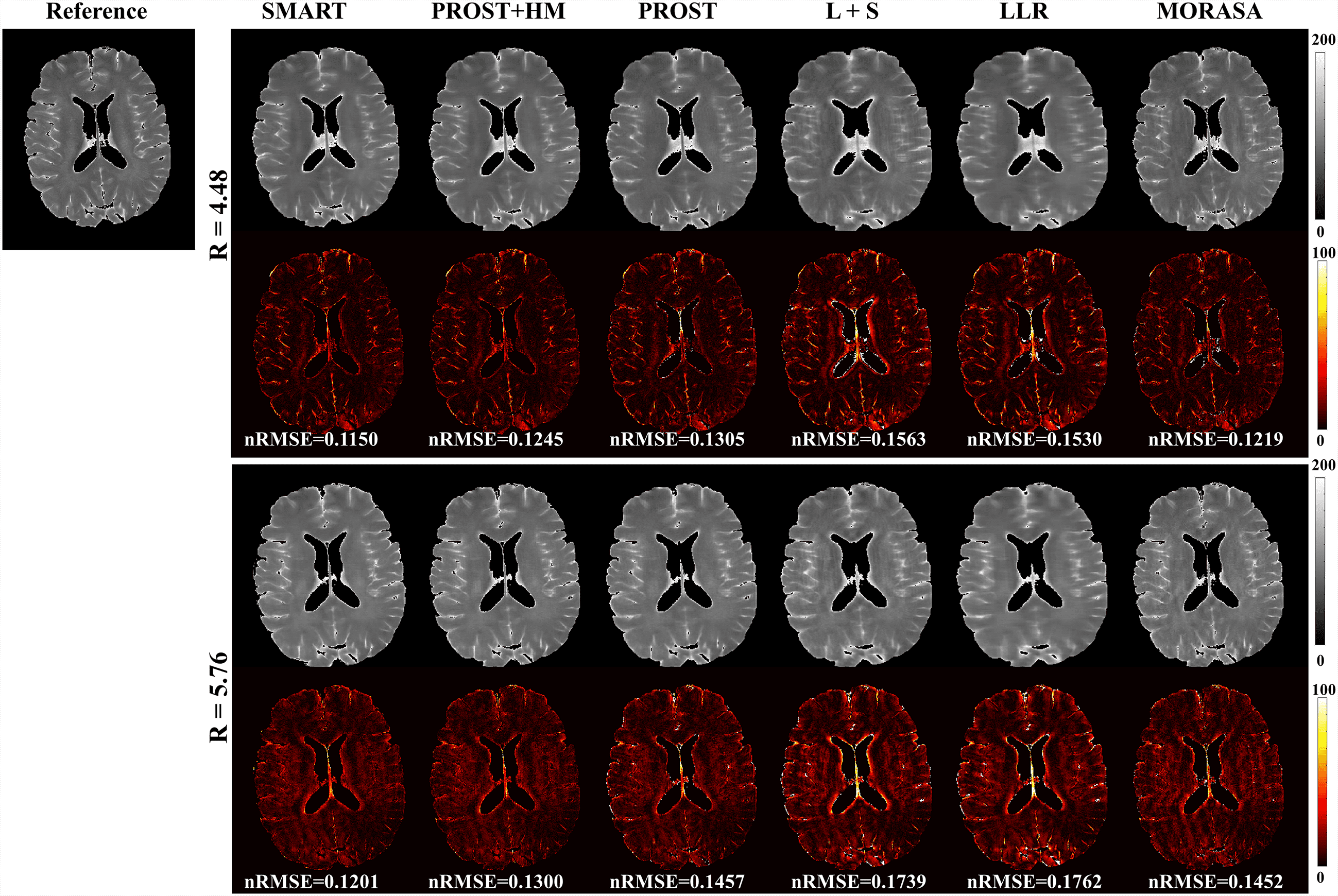}}
\caption {$\text T_{1\rho}$ maps estimated from prospective 2D brain dataset with R = 4.48 and 5.76 using the SMART, PROST + HM, PROST, L + S, LLR, and MORASA methods. The corresponding error images with respect to the reference estimated from fully sampled data are also shown. The quantitative nRMSE results are shown at the bottom of each error image.}
\label{figure S2}
\end{figure*}

Supplementary information Fig.\ref{figure S5} shows the  $\text T_{1\rho}$ maps and the corresponding error maps obtained using the above methods for R = 11.26 and 13.21 in 3D imaging. Similar conclusions can also be drawn as those from Fig. \ref{figure S4}.

\subsection{Parameter Selection}
\par To clearly demonstrate the necessity of updating tissue maps in the iteration, the representative zero-filled reconstruction with R = 6, the intermediate results at the 12\textit{th} iteration , the parameter histogram, the estimated $\text T_{1\rho}$ map, and the corresponding tissue groups and the ranks of tissue groups are included in the Supplemental information Fig. \ref{tissue_clustering}. To show the tissue group, we randomly selected 10 of them and chose 10 pixels from each of the group randomly, since the total tissue groups are too much to be shown. Then, an image block with the size of $10 \times 5$ can be constructed for each tissue group, where 10 is the pixel number chosen and 5 is the TSL number. The image blocks are shown as an example of the tissue groups. To calculate the rank of a tissue group, we construct the low-rank matrix with all pixels in the tissue group. The size of the matrix is $N_p \times 5$ where $N_p$ is the pixel number in the group and 5 is the TSL number. The rank value of the matrix was calculated using the SVD method. The rank curves of all tissue groups are shown.
\subsection{Reconstruction Time}
The reconstruction time for each method used in the 2D (R = 4 ) and 3D (R = 6.76) imaging applications is listed in Supplementary information Table \ref{table S3}.

\subsection{Algorithms}
The algorithms for HOSVD and the SMART method are shown in Algorithms 1 and 2, respectively.

\begin{table*}[!htbp]
  \caption{IMAGING PARAMETERS FOR THE 2D AND 3D $\text T_{1\rho}$ DATASETS USED IN THIS STUDY}
\begin{center}
\setlength{\tabcolsep}{0.58mm}{
\begin{tabular}{c|cccccccc}
\hline \hline \multirow{3}{*}{ Retro } & & Scanner & Coils & Matrix & FOV( $\left.\mathrm{mm}^{2}\right)$ & Thickness $(\mathrm{mm})$ & TSL $(\mathrm{ms})$ & Scan time $(\mathrm{min})$ \\
\cline { 2 - 8 } & 2D & Siemens (phantom) & 12 & $256 \times 256$ & $160 \times 160$ & 5 & $1,20,40,60,80$ & $11.2$ \\
\cline { 2 - 8 } & 2D & Siemens (in vivo) & 12 & $384 \times 384$ & $230 \times 230$ & 5 & $1,20,40,60,80$ & $5.4$ \\
\cline { 2 - 9 } & 3D & UIH & 32 & $240 \times 216 \times 86$ & $240 \times 216$ & 2 & $1,15,25,45,65$ & $49.9$  \\
\hline Pro & 2D & Siemens & 12 & $384 \times 384$ & $230 \times 230$ & 5 & $1,20,40,60,80$ & $1.2(4.48 \times), 0.9(5.76 \times)$  \\
\hline \hline
 \end{tabular}}
\label{table S1}
\end{center}
\end{table*}

\begin{figure*}[!htbp]
\centering{\includegraphics[width=2\columnwidth]{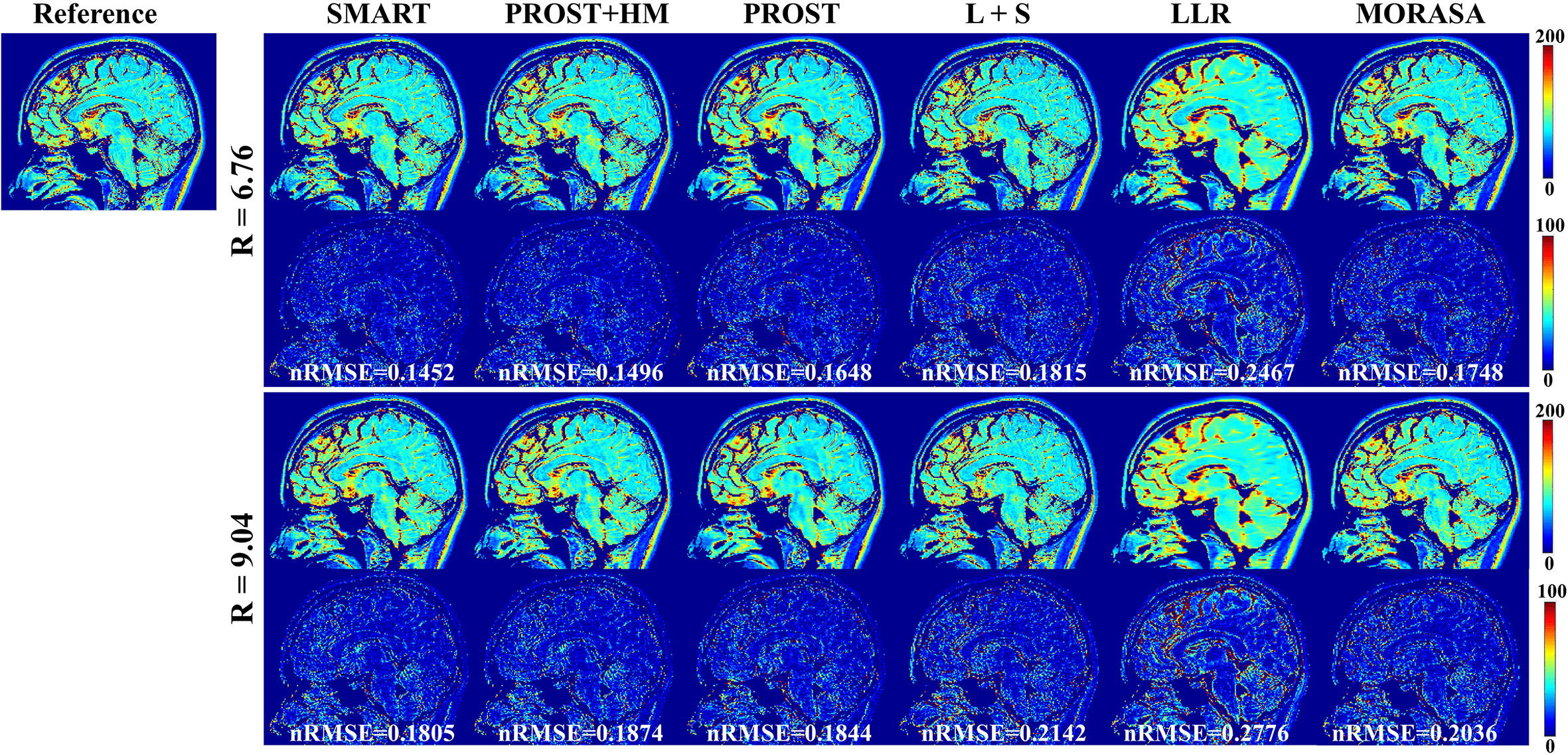}}
\caption{ $\text T_{1\rho}$ maps reconstructed from one slice of the 3D brain dataset with acceleration factors (R) = 6.76 and 9.04 using the SMART, PROST + HM, PROST, L + S, LLR, and MORASA methods. The corresponding error images with respect to the reference estimated from fully sampled data are also shown. The quantitative nRMSE results are shown at the bottom of each error image.}
\label{figure S3}
\end{figure*}

\begin{table*}[!htbp]
  \caption{RECONSTRUCTION PARAMETERS FOR THE 2D AND 3D DATASETS USED IN THIS STUDY}
\begin{center}
\setlength{\tabcolsep}{0.58mm}{
\begin{tabular}{l|ccccccr}
\hline \hline & ADMM iteration & $\lambda_{1}$ & $\lambda_{2}$ & $\mu_{1}$ & $\mu_{2}$ & CG iteration & CG tolerance \\
\hline $2 D$ & 15 & $0.2,0.1,0.1$ & & & & & \\
\cline { 1 - 3 } $3D$ & $15(6.76 \times), 18(9.04 \times)$ & $0.15,0.1,0.1$ & $0.05,0.01,0.01$ & $0.01$ & $0.01$ & 15 & $10^{-7}$ \\
\hline \hline
\end{tabular}}
\label{table S2}
\end{center}
\end{table*}

 \begin{table*}[htbp]
  \caption{RECONSTRUCTION TIME OF EACH METHOD}
\begin{center}
\begin{threeparttable}
{
\begin{tabular}{l|cccccc}
\hline \hline Methods & SMART & PROST+HM & PROST & L+S & LLR & MORASA \\
\hline 2D ($\mathrm{R}=4$) & $369.87$ & $358.25$ &$299.66$ & $437.94$ & $115.46$ & $304.55$ \\
\hline 3D ($\mathrm{R}=6.76$)  & $5981.11$ & $5913.78$ & $5898.54$ & $1197.88$ & $1767.33$& $3579.69$ \\
\hline \hline
\end{tabular} }
\begin{tablenotes}    
        \footnotesize              
      \item  Unit of reconstruction time: second.     
      \end{tablenotes}           
    \end{threeparttable}
\label{table S3}
\end{center}
\end{table*}

\begin{figure*}[!htbp]
\centering{\includegraphics[width=2\columnwidth]{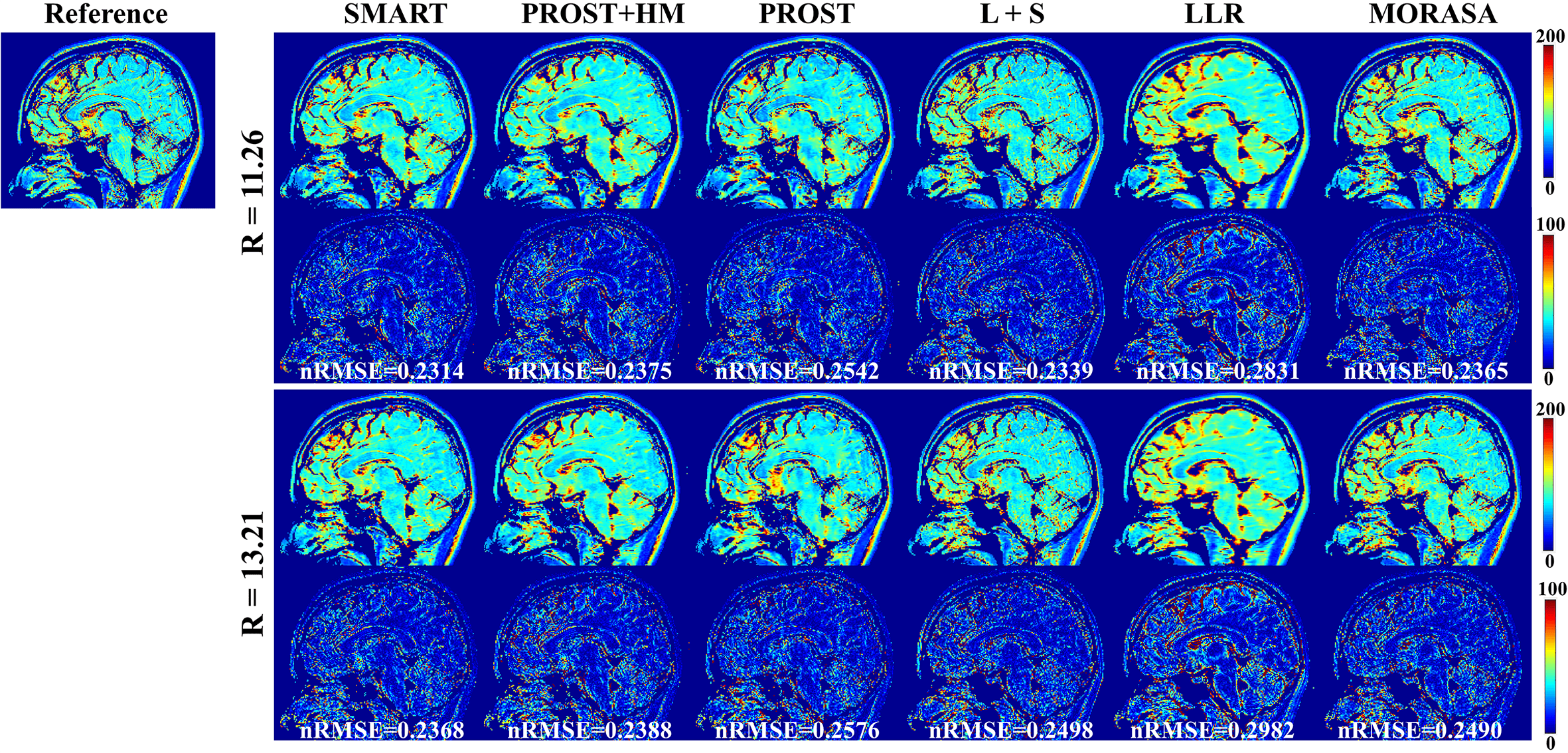}}
\caption{ $\text T_{1\rho}$ maps reconstructed from one slice of the 3D brain dataset with acceleration factors (R) = 11.26 and 13.21 using the SMART, PROST + HM, PROST, L + S, LLR, and MORASA methods. The corresponding error images with respect to the reference estimated from fully sampled data are also shown. The nRMSEs are shown at the bottom of the error images.}
\label{figure S5}
\end{figure*}

\begin{figure*}[!htbp]
\centering{\includegraphics[width=2.1\columnwidth]{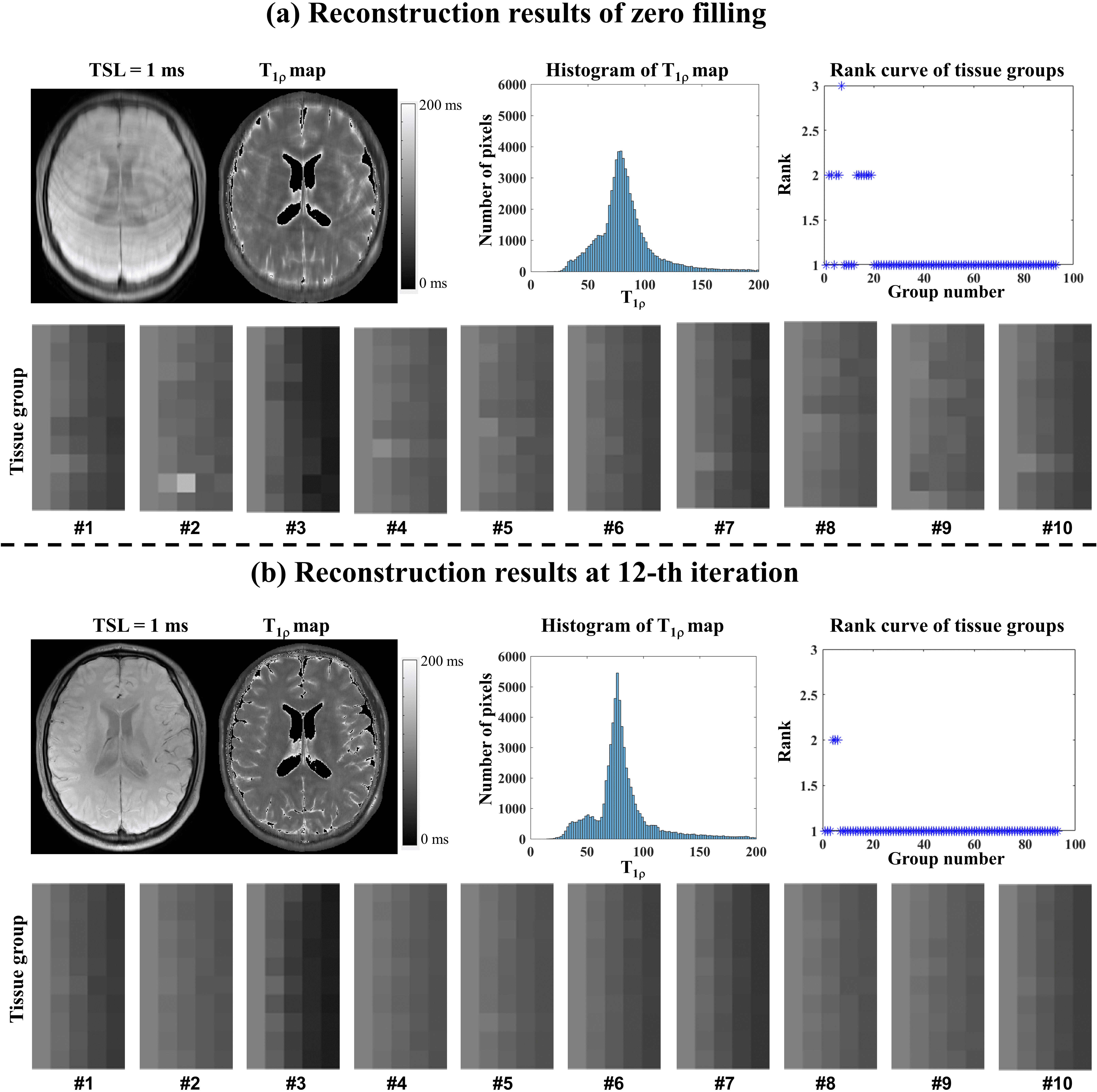}}
\caption{ Zero-filling reconstruction (a) and intermedium reconstruction at the 12\textit{th} iteration (b) results for R = 6. The initial $\text T_{1\rho}$-weighted image of the zero-filling reconstruction shows strong aliasing artifact, and the $\text T_{1\rho}$ map estimated from zero-filling images is quite blurred. The quality of $\text T_{1\rho}$-weighted image and $\text T_{1\rho}$ map was greatly improved at 12\textit{th} iteration. After iteration, the distribution of the parameter histogram also changed (shown in (b)). In the figure of tissue groups, the pixel intensities along pixel number direction are not as uniform as those at 12\textit{th} iteration. The ranks in several groups are reduced from 2 to 1.}
\label{tissue_clustering}
\end{figure*}

\subsection {  Convergence analysis} The convergence of the proposed  method is still an open problem\cite{RN678,RN2005}. Although empirical evidence suggests
that the reconstruction algorithm has very strong convergence
behavior, the satisfying proof is still under study. In this study, we analyze the time complexity of the SMART method using the $\mathcal{O}$ notation, and we give a weak convergence result with the following \textit{Theorem} :
\par \textit{Theorem 1}:
Suppose Assumption (A1) in the Proof section holds, and the multipliers are bounded. Then, we have 
\begin{equation}
\min _{i \in[N]}\left\|\left(\textbf X^*, \mathcal{T}^*, \mathcal{Z}^*, \alpha_{1}^*, \alpha_{2}^*\right)\right\|_{F}^{2} \leq \mathcal{O}\left(\frac{1}{N}\right)
\end{equation}
where $\left(\textbf X^*, \mathcal{T}^*, \mathcal{Z}^*, \alpha_{1}^*, \alpha_{2}^*\right)$ denotes the subgradient of the augmented Lagrangian function $L^i\left(\textbf X, \mathcal{T}, \mathcal{Z}, \alpha_{1}, \alpha_{2}\right)$ and $N$ is the total number of iterations. 
\par The proof of \textit{Theorem} \textit 1 is described in the next section.
\subsection{ Proof}
In this section,  we provide  proofs on the following assumptions:
\par (A1) The sequence of paired proximal operators $(\Gamma^n, \Gamma^{n+1})$  is asymptotically nonexpansive with a sequence \{ $\epsilon^{n+1}$ \}, e.g.,
\begin{equation}
    \|\Gamma^n(u)-\Gamma^{n+1}(u)\|_F^2 \leq (1+\epsilon^{n+1})\|u-v\|_F^2, \forall{u,v,n}
\end{equation}
The assumption (A1) is a standard assumption for analyzing the convergence of the learned iterative algorithms, which also can be found in \cite{RN2008,RN2009}.

Considering the operator $P_i$ is performed at every iteration in,
let $\Gamma =[V(\mathcal{T}_1),\ldots, V(\mathcal{T}_i),\ldots, V(\mathcal{Z}_1),\ldots,V(\mathcal{Z}_j),\ldots]^T$, $\alpha =[V(\alpha_{11}),\ldots, V(\alpha_{1i}),\ldots, V(\alpha_{21}),\ldots,V(\alpha_{2j}),\ldots]^T$,and $Q=[P^n_1,\ldots, P^n_i,\ldots,H_1,\ldots,H_j,\ldots]$, where $V$ denotes an operator which vectorizes the tensor. Then (\ref{Lagrangian model}) can be expressed as
\begin{equation}
\label{new SMART Model with Lagrangian iteratively}
\begin{gathered}
L^n\left(\textbf X, \Gamma, \alpha \right):=\frac{1}{2}\left\|E \textbf X-\textbf Y\right\|_{F}^{2}+\lambda_{1}\sum_{i}
\left\|\mathcal{T}_i\right\|_{*}+
\lambda_{2}\sum_{i}
\left\|\mathcal{Z}_j\right\|_{*}\\
-\left\langle \alpha, \Gamma-Q^n(\textbf X)\right\rangle
+ \frac{\mu}{2}\left \| \Gamma - Q^n(\mathbf X)  \right \|^2_2
\end{gathered}
\end{equation}
 According to the $\frac{\mu}{2}$-strongly convex of $ L^n (\textbf X^n, \Gamma, \alpha^n )$ with respect to $\Gamma$, we have 
\begin{equation}
\label{eq3}
\begin{gathered}
    L^n (\textbf X^n, \Gamma^n, \alpha^n )- L^n (\textbf X^n, \Gamma^{n+1}, \alpha^n )\geq \frac{\mu}{2}\| \Gamma^n - \Gamma^{n+1}\|_2^2
\end{gathered}
\end{equation}
 Since $ L^n (\textbf X, \Gamma^{n+1}, \alpha^n )$ is also $\frac{\mu}{2}$-strongly convex with respect to $\mathbf X$, we have
\begin{equation}
\label{eq4}
\begin{gathered}
      L^n (\mathbf X^n, \Gamma^{n+1}, \alpha^n )- L^n (\mathbf X^{n+1}, \Gamma^{n+1}, \alpha^n ) \\
      \geq \frac{\mu}{2}\| \mathbf X^{n+1} - \mathbf X^{n}\|^2_F
\end{gathered}
\end{equation}
According to the update rule of the multiplier $\alpha$, we have
\begin{equation}
\label{eq5}
\begin{gathered}
      L^n (\mathbf X^{n+1}, \Gamma^{n+1}, \alpha^n )- L^n (\mathbf X^{n+1}, \Gamma^{n+1}, \alpha^{n+1} ) \\
     = -\frac{1}{\mu} \| \alpha^{n+1} - \alpha^{n}\|_2^2
\end{gathered}
\end{equation}
On the other hand, we have
\begin{equation}
\label{eq6}
\begin{aligned}
      L^n (\mathbf X^{n+1}, \Gamma^{n+1}, \alpha^{n+1} )- L^{n+1} (\mathbf X^{n+1}, \Gamma^{n+1}, \alpha^{n+1} )\\
      \geq  -\left\langle\alpha^{n+1},Q^{n+1}(\mathbf X^{n+1})-Q^{n}(\mathbf X^{n+1})\right \rangle\\
      +\frac{\mu}{2}\|Q^{n+1}(\mathbf X^{n+1})-Q^{n}(\mathbf X^{n+1})\|_2^2=0
\end{aligned}
\end{equation}
where the sencond equality comes from the non-expansion assumption (A1) of $Q^n$. With respect to the first-order optimization condition for the update on $\mathbf X$, that is, $\nabla_{\mathbf X} L^n (\mathbf X^{n+1}, \Gamma^{n+1}, \alpha^n)=0$, we have
\begin{equation}
\begin{gathered}
      0= E^*(E\mathbf X^{n+1}-\mathbf Y)+
      (Q^n)^*\alpha^n -
      \mu (Q^n)^*(Q^n(\mathbf X^{n+1}-\Gamma^{n+1}))\\
      =E^*(E\mathbf X^{n+1}-\mathbf Y)+(Q^n)^*\alpha^{n+1}
\end{gathered}
\end{equation}
where $E^*$ and $Q^*$ represent the Hermitian adjoint of operators $E$ and $Q$, respectively. Then we have 
\begin{equation}
\label{eq7}
\begin{aligned}
      \underline{\sigma}_{Q^n} 
      \left \|\alpha^{n+1}-\alpha^n \right\|_2 \leq 
     \|E^*E(\mathbf X^{n+1}-\mathbf X^n)\|_F \\
    \leq \bar{\sigma}_{E^*E} \left \|\mathbf X^{n+1}-\mathbf X^n \right \|_F
\end{aligned}
\end{equation}
where $\underline{\sigma}_{Q^n} $ denotes the minimum singular value of $Q^n$, and $\bar{\sigma}_{E^*E} $ denotes the maximum singular value of $E^*E$.
\par Combining (\ref{eq3}), (\ref{eq4}), (\ref{eq5}), (\ref{eq6}), (\ref{eq7}), we have
\begin{equation}
\label{relation1}
\begin{aligned}
      L^n(\mathbf X^n, \Gamma^n, \alpha^n)-L^{n+1}(\mathbf X^{n+1}, \Gamma^{n+1}, \alpha^{n+1}) \\
    \geq 
      (\frac{\mu}{2}-\frac{\bar{\sigma}^2_{E^*E}}{\underline{\sigma}^2_{Q^n}})\|\mathbf X^{n+1}-\mathbf X^n\|^2_F+\frac{\mu}{2}\|\Gamma^{n+1}-\Gamma^{n}\|^2_2
\end{aligned}
\end{equation}
Summing both sides of the above inequality from 0 to $N$, we have

\begin{equation}
\begin{aligned}
&\frac{N}{2}\left[\left(\frac{\mu}{2}-\frac{\bar{\sigma}_{E^{*} E}^{2}}{\underline{\sigma}_{Q^{n}}^{2}}\right)\left\|\mathbf{X}^{n+1}-\mathbf{X}^{n}\right\|_{F}^{2}+\frac{\mu}{2}\left\|\Gamma^{n+1}-\Gamma^{n}\right\|_{2}^{2}\right]\\
&\leq L^{0}\left(\mathbf{X}^{0}, \Gamma^{0}, \alpha^{0}\right)-L^{N}\left(\mathbf{X}^{N}, \Gamma^{N}, \alpha^{N}\right)<+\infty
\end{aligned}
\end{equation}
On the other hand, let $\left(\mathbf X_{n+1}^{*}, {\Gamma}_{n+1}^{*}, {\alpha}_{n+1}^{*}\right) \in \partial L\left(\mathbf X^{n+1}, {\Gamma}^{n+1}, {\alpha}^{n+1}\right)$, we have 
\begin{equation}
\label{assump1}
\left\{
\begin{array}{l}
{\Gamma}_{n+1}^{*} \in {\alpha}^{n+1}-\mu\left(Q^{n+1}  \mathbf X^{n+1}-{\Gamma}^{n+1}\right)+\lambda \sum_{p} \partial\left\|\Gamma_{p}^{n+1}\right\|_{*} \\
\mathbf X_{n+1}^{*}=E^{*}\left(E \mathbf X^{n+1}-\mathbf Y\right)-Q^{n+1}{\alpha}^{n+1}\\
+\mu {(Q^{n+1})}^{*} \left(Q^{n+1} \mathbf X^{n+1}-{\Gamma}^{n+1}\right) \\

{\alpha}_{n+1}^{*}={\Gamma}^{n+1}-Q^{n+1} \mathbf X^{n+1}
\end{array}\right.
\end{equation}
where $p$ is equal to the sum of $i$ and $j$. According to the iterative rule of ADMM algorithm, we have
\begin{equation}
\label{assump2}
\begin{aligned}
\left\{
\begin{array}{l}
0 \in {\alpha}^{n}-\mu\left(Q^n  \mathbf X^{n}-\Gamma^{n+1}\right)+\lambda  \sum_{p} \partial\left\|\Gamma_p^{n+1}\right\|_* \\
0=  E^*\left(E \mathbf X^{n+1}-\mathbf Y\right)-Q^{n}\alpha^{n}\\
+\mu (Q^{n})^* \left(Q^{n+1} \mathbf X^{n+1}-{\Gamma}^{n+1}\right) \\
0=\alpha_{n+1}-\alpha_n+\mu ({\Gamma}^{n+1}-Q^{n} \mathbf X^{n+1})
\end{array}\right.
\end{aligned}
\end{equation}
Combing (\ref{assump1}) and (\ref{assump2}), we have
\begin{equation}
\left\{\begin{array}{l}
{\Gamma}_{n+1}^{*}={\alpha}^{n}-{\alpha}^{n+1}-\mu  \left(Q^{n+1}\boldsymbol X^{n+1}-Q^n\boldsymbol X^{n}\right) \\
\boldsymbol X_{n+1}^{*}=Q^n{\alpha}^{n}-Q^{n+1}{\alpha}^{n+1} \\
{\alpha}_{n+1}^{*}=\frac{1}{\mu}\left({\alpha}^{n}-{\alpha}^{n+1}\right)
\end{array}\right.
\end{equation}
Then, we have 
\begin{equation}
\begin{aligned}
&\left\|\left({\Gamma}_{n+1}^{*}, \boldsymbol X_{n+1}^{*}, {\alpha}_{n+1}^{*}\right)\right\|_{F}^{2} \\
 \leq &\left(2+(1+\epsilon^n)+\frac{1}{\mu^{2}}\right)\left\|{\alpha}^{n}-{\alpha}^{n+1}\right\|_{2}^{2}\\
& +2 \mu^{2} (1+\epsilon^n)\left\|\boldsymbol X^{n+1}-\boldsymbol X^{n}\right\|_{F}^{2} \\
  \leq &\left[\frac{\bar{\sigma}_{E^{*} E}^{2}}{\underline{\sigma}_{{Q}^n}^{2}}\left(3+\epsilon^n+\frac{1}{\mu^{2}}\right)+2 \mu^{2}  (1+\epsilon^n)\right]\left\|\boldsymbol X^{n+1}-\boldsymbol X^{n}\right\|_{F}^{2}
\end{aligned}
\end{equation}
where ${\bar \sigma}_{Q^n}$ denotes the largest singular values of $Q^n$,  the first inequality is due to the nonexpensive assumption of $Q^n$, and the second inequality is due to
the relation (\ref{eq7}). By the convexity of norm, we have
\begin{equation}
\begin{aligned}
& \min _{i \in[N]}\left\|\left({\Gamma}_{i}^{*}, \boldsymbol X_{i}^{*}, {\alpha}_{i}^{*}\right)\right\|_{F}^{2} \\
\leq 
& \frac{1}{N} \sum_{n=1}^{N}\left\|\left({\Gamma}_{n}^{*}, X_{n}^{*}, {\alpha}_{n}^{*}\right)\right\|_{F}^{2} \\
\leq
& \left[\frac{\bar{\sigma}_{E^{*} E}^{2}}{\underline{\sigma}_{{Q}^n}^{2}}\left(3+\epsilon^n+\frac{1}{\mu^{2}}\right)+2 \mu^{2}  (1+\epsilon^n)\right]\\
& \cdot\frac{1}{N} \sum_{n=1}^{N}\left\|\boldsymbol X^{n+1}-\boldsymbol X^{n}\right\|_{F}^{2}  \\
\leq
& \left[\frac{\bar{\sigma}_{E^{*} E}^{2}}{\underline{\sigma}_{{Q}^n}^{2}}\left(3+\epsilon^n+\frac{1}{\mu^{2}}\right)+2 \mu^{2}  (1+\epsilon^n)\right]\left(\frac{\mu}{2}-\frac{\bar{\sigma}_{E^{*} E}^{2}}{\mu \underline{\sigma}_{{Q^n}}^{2}}\right)^{-1} \\
& \cdot \frac{L\left(\boldsymbol X^{0}, {\Gamma}^{0}, {\alpha}^{0}\right)-L\left(\boldsymbol X^{N}, {\Gamma}^{N}, {\alpha}^{N}\right)}{N} 
\end{aligned}
\end{equation}
From aforementioned analysis, we know that we can choose $\mu=\mathcal{O}( \frac{\bar{\sigma}_{E^{*} E}}{\underline{\sigma}_{Q}})$ which implies that
\begin{equation}
\min _{i \in[N]}\left\|\left({\Gamma}_{i}^{*}, \boldsymbol X_{i}^{*}, {\alpha}_{i}^{*}\right)\right\|_{F}^{2} \leq \mathcal{O}\left(\frac{1}{N}\right)
\end{equation}

\begin{algorithm}[H]
  \caption{Higher-order Singular Value Decomposition (HOSVD) for SMART reconstruction}
  \label{HOSVD algorithm}
  \begin{algorithmic}[1]
    \Inputs{Third-order data tensor $\mathcal{T} \in C^{N_{1} \times N_{2} \times N_{3}}$ with dimensions $\left(N_{1}, N_{2}, N_{3}\right)$ \\ The regularization parameter $\lambda=\left[\lambda_{1}, \lambda_{2}, \lambda_{3}\right]$\\ }\\
    \State \textbf {ALGORITHM:}
   \State \parbox[t]{\dimexpr\linewidth-\algorithmicindent}{(1) Unfold the tensor $\mathcal{T}$ along its single modes:
      \State  $\mathbf{T}_{(1)}$ : reshapes $\mathcal{T}$ into an $N_{1} \times\left(N_{2}\times N_{3}\right)$ complex matrix.
      \State $\mathbf{T}_{(2)}$ : reshapes $\mathcal{T}$ into an $N_{2} \times\left(N_{1}\times N_{3}\right)$ complex matrix.
      \State  $\mathbf{T}_{(3)}$ : reshapes $\mathcal{T}$ into an $N_{3} \times\left(N_{1}\times N_{2}\right)$ complex matrix. \strut}
      \State \parbox[t]{\dimexpr\linewidth-\algorithmicindent}{(2) Compute the complex SVD of $\mathbf{T}_{(\mathrm{n})}(\mathrm{n}=1,2,3)$ and obtain the orthogonal matrices $\mathbf{U}^{(1)}, \mathbf{U}^{(2)}$, and $\mathbf{U}^{(3)}$ from the n-mode signal subspace. \strut}
      \State \parbox[t]{\dimexpr\linewidth-\algorithmicindent}{(3) Compute the complex core tensor $\mathcal{G}$ related by
      \State  $ \qquad \mathcal{G}=\mathcal{T} \times{ }_{1} \mathbf{U}_{(1)}^{H} \times{ }_{2} \mathbf{U}_{(2)}^{H} \times{ }_{3} \mathbf{U}_{(3)}^{H}$
      \State  which is equivalent to its unfolding forms:
      \State  $\qquad \boldsymbol{G}_{(n)}=\mathbf{U}_{(n)}^{H} \mathbf{T}_{(n)}\left[\mathbf{U}_{(i)} \otimes \mathbf{U}_{(j)}\right],  \text { with } 1 \leq n \leq 3 \text       { and } i \neq j \neq \mathrm{n}$
      \State where $\otimes$ denotes the Kronecker product.\strut}
      \State \parbox[t]{\dimexpr\linewidth-\algorithmicindent}{(4) Compute the high-order singular-value truncation (hard thresholding):
      \State   $\qquad \boldsymbol{G}_{(n)}\left(\boldsymbol{G}_{(n)}<\lambda_{n}\right)=0$ \strut}
      \State \parbox[t]{\dimexpr\linewidth-\algorithmicindent}{(5) Construct back the filtered tensor $\mathcal{T}^{\text {denoise }}$ with the $\mathrm{n}$-mode $(\mathrm{n}=1,2,3)$ unfolding matrix      $\mathbf{T}_{(n)}^{\text {denoise }}$ calculated as follows:
      \State $\qquad \mathbf{T}_{(n)}^{\text {denoise }}=\mathbf{U}_{(n)} \mathcal{G}\left[\boldsymbol{U}_{(i)} \otimes \boldsymbol{U}_{(j)}\right]^{H} \text { with } 1 \leq n \leq 3 \text { and } i \neq j \neq \mathrm{n}$ \strut}\\
    \State \textbf {OUTPUT:} {The denoised tensor $\mathcal{T}^{\text {denoise }}$ is obtained by folding.}
  \end{algorithmic}
  \end{algorithm}

\begin{algorithm}[H]
  \caption{Simultaneously spatial patch-based low-rank tensor and parametric group-based low-rank tensor (SMART)}
  \label{SMART algorithm}
  \begin{algorithmic}[1]
    \Inputs{ $\mathbf{Y}$ : undersampled k-space data \\ blocking matching parameters:\\$\lambda_{m}$ : normalized $l_{2}$-norm distance threshold\\ $N_{p, \max }$ : maximum similar patch number\\$b$ : patch size\\reconstruction parameters:\\$A D M M_{\text {iter }}$ :ADMM iterations\\$C G_{\text {iter }}: \mathrm{CG}$ iterations\\$\lambda_{1}=\left[\lambda_{1(1)}, \lambda_{1(2)}, \lambda_{1(3)}\right], \lambda_{2}=\left[\lambda_{2(1)}, \lambda_{2(2)}, \lambda_{2(3)}\right]$: regularization parameters\\$E$ : Encoding operator\\ $\alpha_{1}, \alpha_{2}$ : Lagrangian multipliers\\
    $\mu_{1}, \mu_{2}$ : Penalty parameters}\\
    \State \textbf {ALGORITHM:}
    \State \parbox[t]{\dimexpr\linewidth-\algorithmicindent}{1. Initialize $\mathbf{X}^{1}=E^{*}(\mathbf{Y}), \alpha_{1}^{1}=0, \alpha_{2}^{1}=0, \mu_{1}= 0.01, \mu_{2}=0.01$.\strut}
      \State  {2. Estimate the initial $\mathrm{T}_{1 \rho}$ map $\mathrm{T}_{1 \rho}^{1}$ from $\mathbf{X}^{\mathbf{1}}$ using (16).}\\
     3. \For { $\mathrm{n}=1, \ldots, A D M M_{\text {iter }}$}
      \State \parbox[t]{\dimexpr\linewidth-\algorithmicindent}{[1] Groups of similar spatial patch extraction on $\mathbf{X}^{n}+\frac{\alpha_{1}^{n}}{\mu_{1}}$ centered at spatial location $i$ using the blocking
            matching method with the normalized $l_{2}$-norm distance threshold $\lambda_{m}$, maximum similar patch number $N_{\text {patch,max }}$, and patch size $b$ \strut}
      \State  \parbox[t]{\dimexpr\linewidth-\algorithmicindent}{[2] Spatial tensor construction of $\mathcal{T}_{i}^{n}$ for each group of similar patches \strut}
      \State \parbox[t]{\dimexpr\linewidth-\algorithmicindent}{[3] Solve optimization of problem (P2) in (10): HOSVD-based denoising (see Algorithm 1) with $\lambda_{1}$ \strut}
      \State \parbox[t]{\dimexpr\linewidth-\algorithmicindent}{[4] Clusters of the same tissue extraction from $\mathbf{X}^{n}+\frac{\alpha_{2}^{n}}{\mu_{2}}$ using histogram analysis of $\mathrm{T}_{1 \rho}^{n}$\strut}
               \textbf {Output:} denoised spatial tensor $\mathcal{T}_{i}^{n+1}$ and the estimated image $\widetilde{\mathbf{T}}^{n+1}$ 
      \State  \parbox[t]{\dimexpr\linewidth-\algorithmicindent}{[5] Parametric tensor construction of $\mathcal{Z}_{j}^{n}$ for each cluster of the same tissue \strut}
      \State  \parbox[t]{\dimexpr\linewidth-\algorithmicindent}{[6] Solve optimization of problem (P3) in (12): HOSVD-based denoising (see Algorithm 1) with                $\lambda_{2}$\strut} 
               \textbf {Output: }denoised spatial tensor $\mathcal{Z}_{j}^{n+1}$ and and the estimated image $\tilde{\mathbf{Z}}^{n+1}$ 
      \State \parbox[t]{\dimexpr\linewidth-\algorithmicindent}{[7] Solve optimization of problem (P1) in (9) using the $C G$ algorithm \strut}
            \textbf {Output:} reconstructed images $\mathbf{X}^{n+1}$
      \State \parbox[t]{\dimexpr\linewidth-\algorithmicindent}{[8] Update $\alpha_{1}$ using (13): $\alpha_{1}^{n+1}=\alpha_{1}^{n}+\mu_{1}\left(\mathbf{X}^{n+1}-\widetilde{\mathbf{T}}^{n+1}\right)$ \strut}
      \State \parbox[t]{\dimexpr\linewidth-\algorithmicindent}{
      [9] Update $\alpha_{2}$ using (14): $\alpha_{2}^{n+1}=\alpha_{2}^{n}+\mu_{2}\left(\mathbf{X}^{n+1}-\tilde{\mathbf{Z}}^{n+1}\right)$ \strut}
      \State \parbox[t]{\dimexpr\linewidth-\algorithmicindent}{
      [10] if (n mod 3 = 0 ), estimate the $T_{1 \rho}$ map $\mathrm{T}_{1 \rho}^{n+1}$ from $\mathbf{X}^{n+1}$ \strut}
      \EndFor\\
    \State \textbf {OUTPUT:} {Reconstructed $T_{1 \rho}$-weighted image series $\mathbf{X}$}
  \end{algorithmic}
\end{algorithm}
\end{document}